\documentclass[%
unsortedaddress,
twocolumn,
showkeys,
amsmath,amssymb,
aps, pre, floatfix,
longbibliography,
10pt]{revtex4-2}

\usepackage{graphicx}
\usepackage{dcolumn}
\usepackage{bm}
\usepackage{physics} 
\usepackage{placeins}

\usepackage{hyperref}

\newcommand{\bv}{\mathbf{v}}
\newcommand{\bx}{\mathbf{x}}
\newcommand{\bz}{\mathbf{z}}

\newcommand{\bX}{\mathbf{X}}
\newcommand{\bC}{\mathbf{C}}
\newcommand{\bD}{\mathbf{D}}
\newcommand{\bF}{\mathbf{F}}
\newcommand{\bG}{\mathbf{G}}
\newcommand{\bJ}{\mathbf{J}}
\newcommand{\bL}{\mathbf{L}}
\newcommand{\bT}{\mathbf{T}}
\newcommand{\bI}{\mathbf{I}}
\newcommand{\bR}{\mathbf{R}}

\newcommand{\bLambda}{\mathbf{\Lambda}}

\usepackage[
top=2.5cm,
bottom=2.5cm,
left=2.5cm,
right=2.5cm
]{geometry}

\begin{document}

\title{\textbf{Stochastic thermodynamics and predictability of Ornstein–Uhlenbeck processes: an application to the Madden–Julian Oscillation}}

\author{Roberta Benincasa}
\affiliation{Department of Physics and Astronomy, University of Bologna, Bologna, Italy}
\email{roberta.benincasa2@unibo.it}
\author{Danni Du}
\affiliation{Princeton University, Princeton, NJ, USA}
\author{Gregory S. Duane}
\affiliation{Department of Atmospheric and Oceanic Sciences, University of Colorado, Boulder, CO, USA}
\author{Jeffrey B. Weiss}
\affiliation{Department of Atmospheric and Oceanic Sciences, University of Colorado, Boulder, CO, USA}

\date{September 29, 2026}

\begin{abstract}
Climate oscillations are intrinsically out-of-equilibrium phenomena, yet the connection between their predictability and nonequilibrium thermodynamic properties remains unexplored. We develop a coarse-grained stochastic thermodynamic framework for Ornstein-Uhlenbeck (OU) processes and apply this framework to climate data via linear inverse models (LIMs), linking predictability, entropy production, dynamical activity, and path action. Climate oscillations are often studied in a 2d phase space of climate indices, which are even under time-reversal, motivating our focus on 2d OU dynamics with even variables. We find that coordinate-invariant properties of 2d OU processes can be reduced to a minimal three-parameter description: two diffusivities and a phase-space rotation. We show that these quantities fundamentally characterize distinct aspects of the dynamics. We find that higher irreversibility and dynamical activity do not necessarily imply reduced predictability. Rather, it is the path action which is most closely connected to a commonly used measure of predictability, the anomaly correlation coefficient (ACC). As an illustrative example, we apply this framework to the Madden-Julian Oscillation (MJO) over the twentieth century.  We find that the MJO has become simultaneously more predictable, as seen previously, while at the same time becoming more irreversible and more active. Predictability is driven predominantly by the reduced mean diffusion. Entropy production reflects the combined evolution of all three parameters. These results suggest that stochastic thermodynamics provides a powerful lens for studying nonequilibrium climate phenomena.


\end{abstract}

\maketitle

\section{Introduction}
\label{sec:intro}

Climate is fundamentally a nonequilibrium system driven by the coupling between two thermal reservoirs: the Sun at $6000$ K and the cosmic background radiation at $3$ K. Climate oscillations are dominant coherent modes of natural variability which emerge as fluctuations within the nonequilibrium steady state of the climate system. They are typically characterized using a small number of indices, constructed from combinations of climate variables that capture the dominant features of the mode, yielding a compact phase-space representation of the dynamics. The resulting phase-space trajectories exhibit rotational probability currents and irreversible behavior, the hallmarks of nonequilibrium dynamics. One prominent example is the Madden-Julian Oscillation (MJO), an eastward-propagating pattern of atmospheric convection, winds, and pressure. The pattern travels eastward through the tropics with a 30-60-day timescale, and represents one of the leading modes of intraseasonal climate variability. We focus here on climate oscillations in a steady climate. Anthropogenic climate change modifies the climate on decadal timescales, and the 30-60 day MJO timescale gives a clear timescale separation. We thus treat climate change as a succession of steady-states.

The predictability of chaotic systems is commonly divided into two types \cite{lorenz1975climatic}: the initial-value problem, or \textit{predictability of the first kind}, and the boundary-value problem, or \textit{predictability of the second kind}. While the latter concerns the response to changes in boundary conditions, such as external forcing, the former—our focus here—concerns the growth of initial-condition errors and the resulting loss of predictability. Multiscale complex systems can retain statistical initial-value predictability beyond the detailed predictability lost in fast chaos. In climate, for example, weather-chaos destroys the predictability of daily precipitation after 1-2 weeks, and beyond that timescale climate oscillations emerge from the weather-chaos providing forecast skill on the oscillation timescale.\cite{palmer2006predictability, robertson2025subseasonal, guan2024atmospheric}. Ensemble forecasts provide a natural framework for assessing such predictability by characterizing the range of possible trajectories, and are widely used to assess forecast skill and the limits of predictability. Accurate prediction of climate oscillations can contribute substantially to extended-range forecast skill, yet they remain challenging despite decades of research \cite{subramanian2011madden, ajayamohan2013realistic, bellenger2014enso, chang2015madden, chen2017enso}.

Given the nonequilibrium nature of climate oscillations, a possible approach to understanding and improving their predictability is to study them within the framework of nonequilibrium thermodynamics \cite{weiss2007fluctuation,weiss2009nonequilibrium,weiss2020nonequilibrium}. The mathematical framework we use is the multivariate Ornstein-Uhlenbeck (OU) stochastic differential equation \cite{uhlenbeck1930theory}. OU dynamics is among the simplest nonequilibrium stochastic dynamical systems: the deterministic dynamics is linear, while the stochastic forcing is additive Gaussian white noise. Conveniently, climate science commonly uses the Linear Inverse Modeling (LIM) algorithm to parameterize climate oscillations by OU dynamics. LIM fits OU model parameters to climate oscillation trajectories in a reduced phase space, matching the OU phase space. Trajectories are commonly obtained from both Earth observations and from models across the climate model hierarchy, including the most complex Earth System Models. After several decades of research, it is now well-established that, despite the high dimensional phase space of active climate dynamics, and despite the simplicity of OU dynamics, LIM parameterized OU dynamics is a skillful low-dimensional representation of large-scale climate variability \cite{penland1993prediction,penland1998prediction,newman2003study,alexander2008forecasting,hawkins2009decadal,stevenson2013generalized,cavanaugh2015skill,dias2019remote,lou2020linear,kido2023understanding}. LIM thus provides a natural framework to apply stochastic thermodynamics to climate phenomena \cite{nicolis2017stochastic}. While considerable attention has been devoted to the forecasting skill of LIMs, their potential for elucidating the fundamental dynamics underlying climate oscillations has received comparatively little attention \cite{newman2009important,marsico2026modal}.

In the last decades, stochastic thermodynamics has emerged as a powerful framework for describing the behavior of fluctuating out-of-equilibrium systems with few degrees of freedom. Stochastic thermodynamics extends classical thermodynamic concepts such as work, heat, and entropy to individual stochastic trajectories  \cite{evans1993probability,sevick2008fluctuation,seifert2008stochastic,seifert2012stochastic,shiraishi2023introduction}. Although analyzing the full time-dependent behavior of such far-from-equilibrium systems remains challenging, their stationary states have been extensively studied \cite{seifert2019stochastic}. 

Equilibrium steady-states are time-reversal invariant and exhibit no mean phase space rotation. Nonequilibrium steady-states (NESS's) are characterized by persistent phase space probability currents, trajectory phase space rotation, and broken time-reversal symmetry \cite{hill1966studies, seifert2019stochastic, hatanosasa2001}. OU processes are among the simplest systems capable of exhibiting such behavior, and their linear Gaussian structure allows many thermodynamic quantities to be derived analytically \cite{weiss2007fluctuation,seifert2005entropy,andrieux2007entropy,nicolis2017stochastic}. As a result, OU processes have become a paradigmatic framework for the study of stochastic nonequilibrium dynamics. In their application to complex physical systems, such as the climate system, OU dynamics models the interaction between slowly evolving large-scale variability and rapidly varying unresolved processes represented by stochastic forcing \cite{hasselmann1976stochastic, frankignoul1977stochastic, franzkeokane2017}. 

While the thermodynamics of continuous OU processes is well established \cite{baiesi2009nonequilibrium,baiesi2010nonequilibrium}, its extension to discrete-time observations has been less extensively explored, particularly in the climate literature. This framework was first applied in \cite{weiss2007fluctuation, weiss2009nonequilibrium}, where it was demonstrated that, on monthly timescales, the seasonally averaged tropical sea surface temperature (SST) exhibits finite-time negative entropy production fluctuations. This confirms that, on these timescales, tropical SST variability is a manifestation of fluctuations in a small, fast NESS. These results raise the broader question of whether the predictability of climate oscillations may itself be related to their nonequilibrium thermodynamic properties. A similar question was explored in \cite{vannitsem2023weather}, which found that observations and climate-change projections exhibit decreasing predictability alongside increasing irreversibility of weather-pattern sequences over western Europe, raising the possibility of a connection between the two. 

The aim of this work is to investigate whether there is a connection between the predictability of climate oscillations and selected thermodynamic properties: stochastic entropy production (E), dynamical activity (A), or frenesy, and the path action (A), which we will refer to collectively as EAA \cite{seifert2005entropy, seifert2012stochastic, maes2020frenesy}. To this end, we develop an explicit coarse-grained stochastic thermodynamic framework for overdamped Langevin systems. We apply the framework to the Madden–Julian Oscillation as a test case. Using a reduced coordinate representation, we derive analytical expressions for predictability and EAA, identify distinct rotational and diffusive dynamical regimes, and measure how mean diffusion, diffusive anisotropy, and phase space rotation contribute to the predictability of the MJO over the twentieth century. 

The paper is organized as follows. Section \ref{sec:ou_dynamics} introduces Ornstein–Uhlenbeck processes and the specific coordinate system we use. Section \ref{sec:stoch_thermo} outlines the essential concepts of stochastic thermodynamics, illustrates the coarse-graining method, and connects stochastic thermodynamic EAA to information theory. Section \ref{sec:predictability} introduces the concept of potential predictability as used in climate science. Section \ref{sec:rot_diff_regimes} presents 2d reduced parameter space and identifies different dynamical regimes within the parameter space. The framework is then applied in Section \ref{sec:MJO} to analyze the predictability of the MJO through its LIM OU parameterization. The paper concludes with a discussion of the main results and perspectives (Section \ref{sec:conclusion}).

\section{Ornstein-Uhlenbeck dynamics}
\label{sec:ou_dynamics}

\subsection{Representations}
\label{subsec:representations}

One of the simplest types of climate models is a multivariate stochastic differential equation with linear deterministic damping and additive Gaussian white noise. This model, an overdamped Langevin system, is known as the Ornstein-Uhlenbeck (OU) equation \cite{uhlenbeck1930theory}. Throughout this paper \textit{OU} refers specifically to the overdamped, all-even-parity subclass of Ornstein-Uhlenbeck processes appropriate to position-like climate indices, as opposed to the broader class that can include odd-parity (e.g., velocity-like) variables. The system's state space is described by a real $N$-dimensional state vector, $\bx(t)$. The dynamics are defined by real, constant $N\times N$ matrices, with associated constraints that ensure the dynamics are physically realizable. OU dynamics can be defined in several different representations, all describing exactly the same dynamical system. Here we consider three different representations of OU dynamics, a dynamical representation, an observational representation, and a thermodynamic representation. The defining matrices in each representation are related to the matrices in the other representations by nonlinear matrix relationships.

The OU stochastic differential equation is defined in the \textit{dynamical} $(\bF, \bD)$ representation by two matrices, $\bF$ and $\bD$. The dynamical equation is
\begin{equation}
    \frac{d\bx(t)}{dt} = \bF \bx(t)+\sqrt{2\mathbf{D}}\,\boldsymbol{\xi}(t).
    \label{eq:langevin}
\end{equation}
Here $\bF$ is the drift matrix whose eigenvalues must have negative real parts to ensure stationarity, $\mathbf{D}$ is a symmetric positive-definite diffusion matrix, and $\boldsymbol{\xi}$ is identity-covariance Gaussian white noise. The solution to Eq.~\eqref{eq:langevin} is
\begin{equation}
    \mathbf{x}(t) = \bG(t)\mathbf{x}_0 + \sqrt{2\mathbf{D}}\int_0^t ds \, \bG(t-s)\boldsymbol{\xi}(s),
    \label{eq:ou_solution}
\end{equation}
where $\bG(t) = \exp(\bF t)$ is the deterministic Green function.
The solution has a Gaussian steady-state pdf $P_0(\bx)$ with covariance $\bC_0$ which is obtained from $(\bF,\bD)$ by the Lyapunov equation
\begin{align}
    \bC_0 \bF^T + \bF \bC_0+2 \bD = \mathbf{0}.
    \label{eq:lyapunov}
\end{align}
The associated Fokker-Planck equation has a phase space current $\bJ$ which can be defined in terms of a velocity $\bv$ and the steady-state distribution,
\begin{align}
    \bJ &= \bv P_0(\bx),\\
    \bv &= \left(\bF + \bD \bC_0^{-1}\right)\bx = \mathbf{\Omega} \bx,
\end{align}
where $\mathbf{\Omega}$ is a phase-space frequency.

The LIM algorithm fits OU dynamics to discrete time observations of a trajectory $\bx(t_n)$. The choice of climate indices defining the phase space $\bx$, and the observation timestep $\Delta t$ are crucial subjective choices in the LIM algorithm. It is found empirically that for appropriate choices, LIM produces OU systems that skillfully model climate phenomena. We choose a unit of time such that $\Delta t = 1$ and the observations are then at $t_n = t_0 + n$. For the MJO, we will follow the common choice of $\Delta t = 1$ day, and time is measured in days.

The \textit{observational} $(\bC_0,\bC_{lag-1})$ representation is defined by the equal-time and lag-1 covariance matrices $\bC_0$ and $\bC_{lag-1}$. These are obtained from the observed trajectory by time averages $\langle \ldots \rangle_t$
\begin{align}
    \bC_0 &= \langle \bx(t)\bx^T(t)\rangle_t,\\
     \bC_{lag-1} &= \langle \bx(t+1)\bx^T(t)\rangle_t.
\end{align}
By construction, $\bC_0$ is positive definite and, if the trajectory is a realization of OU dynamics, Eq. \eqref{eq:ou_solution}, is equivalent to the steady-state covariance in Eq. \eqref{eq:lyapunov}. The LIM algorithm calculates  the dynamical representation from the observational representation, using
\begin{equation}
\label{eq:lag1relation}
    \bC_{lag-1} = e^{\bF}\bC_0
\end{equation}
to determine $\bF$, and Eq. \eqref{eq:lyapunov} for $\bD$.

It is often convenient to work in yet another representation, a \textit{thermodynamic} $(\bC_0,\bD,\bL)$ representation, that makes explicit the phase space angular momentum matrix $\bL$ \cite{weiss2020nonequilibrium,Strang2024}. This representation is defined by the three matrices, $\bC_0$, $\bD$, and $\bL$, where 
\begin{align}
    \bL = -\bF \bC_0 -\bD = -\mathbf{\Omega} \bC_0^{-1}.
    \label{eq:L_definition}
\end{align}
showing the relationship between the phase space angular momentum and rotation frequency $\mathbf{\Omega}$. $\bL$ is an antisymmetric matrix, as seen in the Sylvester equation:
\begin{align}
     \bF \bC_0 - \bC_0 \bF^T + 2 \bL = \mathbf{0}.
    \label{eq:Sylvester}
\end{align}
It has units of length$^2$/time, the same units as diffusion and angular momentum. It is also, dimensionally, an area production rate, and quantifies the stochastic area production rate \cite{Strang2024}. Note that $\bD$ and $\bL$ are related to $\bC_0$ and $\bF$ through Eqs.~(\ref{eq:lyapunov}) and (\ref{eq:Sylvester}), respectively, in which the remaining terms differ only in a relative minus sign. The dynamical representation highlights the deterministic drift matrix $\bF$, and $\bC_0$ is defined by the Lyapunov equation Eq.~(\ref{eq:lyapunov}). In the thermodynamic representation, the drift matrix is no longer fundamental and is defined by the Sylvester equation (Eq.~(\ref{eq:Sylvester})). The specified diffusion requires damping in $\bF$ to keep the covariance finite. 

OU models describe both equilibrium and nonequilibrium systems depending on whether the dynamics satisfies or violates, respectively, the condition of detailed balance. The precise form of the detailed-balance condition depends on the time-reversal parity of the state variables. In the general linear Langevin system, time reversal is characterized by a diagonal parity matrix $\boldsymbol{\epsilon}$, whose elements are $\epsilon_i=+1$ for variables that are even under time reversal and $\epsilon_i=-1$ for variables that are odd. The corresponding Onsager--Casimir relation \cite{gardiner2009stochastic} can be written as
\begin{equation}
\boldsymbol{\epsilon}(\bF\bC_0)\boldsymbol{\epsilon}
=(\bF\bC_0)^T.
\end{equation}
Here, we consider an overdamped Langevin system in which all state variables are treated as even under time reversal and assume $\boldsymbol{\epsilon}=\mathbf{I}$. This aligns with the usual interpretation of climate indices as state coordinates of the phenomenon of interest. In the dynamical representation, the Onsager--Casimir relation then reduces to the detailed-balance condition
\begin{equation}
\bF\bD=\bD\bF^T,
\end{equation}
or, equivalently, to $\bL=\bL^T$ in the thermodynamic representation. Since $\bL$ is antisymmetric by construction, detailed balance therefore requires $\bL=0$, whereas $\bL\neq0$ characterizes a nonequilibrium steady state (NESS). Thus, within the class of OU systems considered here, the phase-space angular momentum matrix $\bL$ directly measures the violation of detailed balance and provides a measure of nonequilibrium dynamics. Consequently, for any pair $\bC_0$ and $\bD$, there is a unique equilibrium steady state with $\bL=0$, while different choices of nonzero $\bL$ generate a family of NESSs with the same $\bC_0$ and $\bD$. Detailed balance is also equivalent to the existence of a coordinate transformation that simultaneously diagonalizes $\bF$ and $\bD$, resulting in $N$ uncoupled one-dimensional OU processes. Such a transformation is not possible for a NESS, for which the dynamics is intrinsically multidimensional and exhibits nonzero phase-space rotation. Nonequilibrium complexity in OU systems therefore arises not from nonlinear dynamics, but from the noncommutativity of the dynamical matrices. Note that the result would be the same with all state variables odd, i.e., $\boldsymbol{\epsilon}=-\mathbf{I}$.

The identification of $\bL\neq0$ with nonequilibrium is specific to the present case of same-parity variables: when variables of different time-reversal parity are present, detailed balance may be satisfied for $\bL\neq0$, and $\bL$ is therefore no longer, by itself, a measure of nonequilibrium. 

We consider prediction in the context of a perfect observation ensemble forecasting system: an ensemble of trajectories initialized at $t=0$ with the same $\bx_0$. Perfect observations give a $\delta$-function initial pdf with zero covariance. As the ensemble evolves in time, ensemble members diverge and the ensemble covariance grows. The time-dependent ensemble mean and covariance are
\begin{align}
    \langle \bx(t)\rangle_{ens} &= e^{\bF t}\bx_0 ,\\
    \mathbf{C}_t &= \langle \Delta\mathbf{x}(t) \Delta\mathbf{x}^T(t) \rangle_{ens}\\
        &= 2\int_0^t ds \, e^{\mathbf{F}(t-s)}\mathbf{D}e^{\mathbf{F}^T(t-s)},
    \label{eq:time_cov_matrix}
\end{align}
where $\Delta \bx(t) = \bx(t) -e^{\bF t}\bx_0$ is the departure of an ensemble member from the ensemble mean. As $t \to\infty$, the mean decays to zero and the covariance asymptotes to the steady-state covariance $\bC_0$. The time-dependent covariance matrix can be written in terms of $\mathbf{C}_0$ and $\mathbf{F}$ \cite{weiss2007fluctuation} which explicitly shows the evolution from $\mathbf{0}$ to $\bC_0$:
\begin{equation}
    \mathbf{C}_t = \mathbf{C}_0 - e^{\mathbf{F}t}\mathbf{C}_0 e^{\mathbf{F}^Tt}.
    \label{eq:time_cov_matrix_2}
\end{equation}

The thermodynamic $(\bC_0,\bD,\bL)$ representation allows the OU parameters to be categorized as coordinate-defining parameters and essential dynamical parameters. In every representation, an $N$d OU system is defined by $(3 N^2+N)/2$ independent parameters plus constraints to ensure physicality. Consider invertible coordinate transformations $\bT$, $\bx'=\bT \bx$. The $(\bC_0,\bD,\bL)$ representation has two positive definite matrices, $\bC_0$ and $\bD$, implying there is a $\bT$ that gives $\bC'_0 = \bI$ and $\bD' = \boldsymbol{\bLambda}_D$, where $\bI$ is the identity matrix and $\boldsymbol{\bLambda}_D$ is a diagonal matrix. The coordinate transformation accounts for $N^2$ parameters, leaving $(N^2+N)/2$ dynamical parameters. We will refer to this coordinate system as \textit{reduced} $(\bC_0,\bD,\bL)$ coordinates or $(\bI,\bLambda_\bD,\bL)$ coordinates, and drop the primes. Coordinate-invariant properties only depend on the dynamical parameters, while coordinate dependent properties can be obtained from properties in $(\bI,\bLambda_\bD,\bL)$ coordinates through the usual coordinate transformation mathematics. Coordinate invariant properties in OU dynamics are thus determined by $N$ diffusions and $(N^2-N)/2$ rotations. Each direction in $(\bI, \bLambda_\bD, \bL)$ phase space has a diffusion parameter and each 2d subplane of the phase space has a rotation parameter.

\subsection{\texorpdfstring{$N=2$ Reduced $(\bC_0,\bD,\bL)$ coordinates}
{N=2 Reduced (C0, D, L) coordinates}}
\label{subsec:cdl_coord}

Here we will work in two dimensions, $N=2$, where there are seven parameters in total, four that define the coordinate system, and three - two diffusions and a rotation - that define the dynamics. In  $(\bI,\bLambda_\bD,\bL)$ coordinates, the three defining matrices are
\begin{align}
\bC_0 &= \bI,\\
\bD &= \bLambda_\bD = 
\begin{bmatrix}
\bar d + \Delta d & 0 \\
0 & \bar d - \Delta d
\end{bmatrix},\\
\mathbf{L} &=
\begin{bmatrix}
0 \, & \ell  \\
-\ell \, & 0
\end{bmatrix},
\label{eq:red_cdl_coord}
\end{align}
where $\bar d$ is the mean diffusion strength, $\Delta d$ the diffusion anisotropy, and $\ell$ the rotational probability current. The sign of $\ell$ in $\bL$ is chosen so that $\ell > 0$ corresponds to rotation in the counterclockwise direction. Note that since the transformation to $(\bI,\bLambda_\bD,\bL)$ coordinates is typically non-orthogonal, the diagonal elements of $\bD$ are not the eigenvalues of $\bD$ in the original coordinates. 

In $(\bI,\bLambda_\bD,\bL)$ coordinates, the drift matrix, $\bF = (-\bL - \bD)\bC_0^{-1}$, is 
\begin{align}
\bF &= 
\begin{bmatrix}
-\bar d - \Delta d & \ell \\
-\ell & -\bar d + \Delta d
\end{bmatrix},
\label{eq:drift_red_cdl_coord}
\end{align}
and has eigenvalues $\mu_\bF = -\bar{d} \pm i \omega$ where $\omega = \sqrt{\ell^2-\Delta d^2}$.

\section{Stochastic Thermodynamics}
\label{sec:stoch_thermo}

\subsection{Entropy production, dynamical activity, and path action}
\label{subsec:entropyfrenesyaction}

Stochastic thermodynamics is framed in terms of the trajectory probability. Denote a continuous trajectory segment from $t_0$ to $t_1$ as $\bX = \{\bx(t): t_0 \le t \le t_1\}$. The probability of finding a trajectory segment is $P[\bX]$. The probability depends on the dynamics, of course, and, when needed, we shall include the dynamics in the definition of $P$.

We focus here on three measures of nonequilibrium steady states: the stochastic entropy production, $\sigma$, the dynamical activity, or frenesy, $\phi$, and the path action, $\alpha$. We will refer to this trio of properties as EAA, entropy production, activity, and action.

\textit{Entropy production} is a fundamental quantity in stochastic thermodynamics. It is a measure of irreversibility and an intrinsic indicator of the departure of the system from equilibrium. Consider a trajectory segment $\bX$ with probability $P[\bX]$ and the probability of its time-reversed trajectory $\hat{\bX}$ under time-reversed dynamics $\hat{P}[\hat{\bX}]$. The stochastic entropy production is defined as the logarithm of the ratio of the probabilities of finding the forward and reverse trajectories \cite{seifert2005entropy,seifert2008stochastic,van2015ensemble}:
\begin{equation}
    \sigma[\bX] = \ln\frac{P[\mathbf{X}]}{\hat{P}[\mathbf{\hat X}]}.
    \label{eq:entropy_production_seifert}
\end{equation}
Under time-reversed dynamics, $\bX \leftrightarrow \hat\bX$, $P \leftrightarrow \hat{P}$ and $\sigma \leftrightarrow -\sigma$. In a steady-state, the forward and backward processes are the same, i.e., $P = \hat P$.

The time-symmetric component $\phi$, introduced by Maes \cite{basu2015nonequilibrium,maes2020frenesy} and called \textit{frenesy} or \textit{dynamical activity}, can be written as: 
\begin{equation}
    \phi[\bX] = -\frac{1}{2} \ln(P[\mathbf{X}]\hat{P}[\mathbf{\hat X}])
    \label{eq:frenesy_definition}
\end{equation}
Dynamical activity has emerged as a key quantity in nonequilibrium statistical mechanics, where it complements entropy production by capturing how actively a system explores its state space and contributes to nonequilibrium response and fluctuation properties. It has been identified as a measure of the kinetic aspects of the dynamics. It is associated, for instance, with transition frequencies and escape rates in jump processes. In contrast to entropy production, which identically vanishes at equilibrium, frenesy goes to a nonzero constant. 

By analogy with the Onsager–Machlup action of a stochastic trajectory \cite{onsager1953fluctuations}, the action is
\begin{equation}
    \alpha[\bX] = -\ln P[\bX].
\end{equation}
The EAA are related by
\begin{align}
    \alpha = \phi - \frac{1}{2}\sigma,
\end{align}
expressing the activity and entropy production as the time-symmetric and time-antisymmetric parts of the action, respectively. It is sometimes useful to compare EAA to a given reference process $P_{ref}[\bX]$, and we define $\Delta \sigma= \sigma_P - \sigma_{P_{ref}}$, and similarly for $\phi$ and $\alpha$.

\subsection{Coarse-grained endpoint Ornstein-Uhlenbeck EAA}
\label{subsec:Coarse-graining}

Climate data are available at discrete observation times. Predictability and forecast skill are often framed in terms of the probability of the system being in a state $x_1$ at some fixed time $t_1 = t_0 + t_f$, where $t_0$ is the \textit{initialization time} and $t_f$ is the \textit{forecast lead time}. This probability averages over all possible intervening trajectories. For these reasons, we explore coarse-grained versions of EAA where we average over trajectories with fixed endpoints. Our ultimate focus is on the average predictability of climate phenomena. We shall accordingly also average over initial conditions chosen from the steady-state distribution $P_0(\bx)$. The variation of predictability with different initial conditions is, however, an important topic that could be explored by considering the predictability and thermodynamics as a function of the initial condition. We leave that question to future work.

Consider a trajectory segment $\bX$ from $t_0$ to $t_1$ conditioned on $\bx(t_0)=\bx_0$ and $\bx(t_1) = \bx_1$ with forecast lead time $t_f = t_1 - t_0$, denoted by $\bX^e(\bx_0, \bx_1, t_f)$. The probability of observing an arbitrary trajectory segment $\bX$ can be decomposed as
\begin{equation}
    P[\bX] = P^{\bX^e}[\bX]P^e(\bx_0, \bx_1),
    \label{eq:decomposition_forward_path_prob}
\end{equation}
where $P^{\bX^e}[\bX] \equiv P[\bX|\bx_0,\bx_1]$ is the conditional probability of observing the fixed-endpoint path $\bX^e$, and $P^e(\bx_0, \bx_1)$ is the joint probability of observing the endpoints. The endpoint probability can be further decomposed as
\begin{equation}
    P^e(\bx_0,\bx_1) = P(\bx_1 \, | \, \bx_0)P_0(\bx_0)
    \label{eq:endpoint_prob_decomposition}
\end{equation}
where the first term is the transition probability of going from $\bx_0$ to $\bx_1$ and the second is the steady-state probability of finding the system at $\bx_0$.

We partition the average over all trajectories, $\langle \cdot \rangle_\bX$, into two successive averages: an average over trajectories with fixed endpoints, $\langle \cdot \rangle_{\bX^e}$, followed by an average over the endpoints, $\langle \cdot \rangle_e$
\begin{widetext}
\begin{align}
    \langle F[\bX] \rangle_{\bX} &= \int_{\bX} \mathcal{D} \bX \, F[\bX] P[\bX], \\
    &= \int d\bx_0 d\bx_1  \left(\int_{\bX^e} \mathcal{D} \bX \, F[\bX] P^{\bX^e}[\bX]\right) P^e(\bx_0, \bx_1),\\
    &= \langle \langle F[\bX] \rangle_{\bX^e} \rangle_{e},\\
    &= \langle F^e(\bx_0,\bx_1) \rangle_e,
\end{align}
\end{widetext}
defining endpoint quantities $F^e(\bx_0,\bx_1)  = \langle F[\bX] \rangle_{\bX^e}$.  Similarly, we can define the path probability of the time reversed trajectory $\hat \bX$ as:
\begin{equation}
    P[\hat \bX] = P^{\hat \bX^e}[\hat \bX]P^e(\bx_1,\bx_0).
    \label{eq:decomposition_backward_path_prob}
\end{equation}
where $P^{\hat \bX^e}[\hat \bX] \equiv P[\hat \bX|\bx_1,\bx_0]$ and $\hat \bX^e(\bx_1,\bx_0,t_f)$ is the backward trajectory segment with fixed endpoints. Thus, in the endpoint formulation, the time-reversed trajectory corresponds to the trajectory segment with reversed endpoints. For an OU process, an equivalent procedure would be to consider the backward trajectory as the one obtained from a dynamics with $\bL \rightarrow \bL^T = -\bL$, thus evolving with opposite rotation \cite{da2023entropy}. 

Climate predictability studies often compare the skill of a forecast system against the skill of choosing a forecast randomly from the steady state. This motivates choosing a reference climatological endpoint process: $P_{ref}^e(\bx_0,\bx_1) = P_0^e(\bx_0,\bx_1) = P_0(\bx_0)P_0(\bx_1)$, where subscript $0$ refers to the steady-state of the climatological process.

By analogy with continuous trajectories, we define the endpoint EAA as:
\begin{align}
    \sigma^e(\bx_0,\bx_1) &= \ln\frac{P^e(\bx_0, \bx_1)}{P^e(\bx_1, \bx_0)} ,
    \label{eq:coarse_grained_sigma} \\
    \phi^e(\bx_0,\bx_1) &= -\frac{1}{2}\ln[P^e(\bx_0, \bx_1)P^e(\bx_1, \bx_0)] ,
    \label{eq:coarse_grained_phi} \\
    \alpha^e(\bx_0,\bx_1) &= -\ln P^e(\bx_0, \bx_1),\\
    &=  \phi^e - \frac{1}{2}\sigma^e,
    \label{eq:coarse_grained_alpha}
\end{align}
and define their reference quantities using $P^e_0$. 
Note that because of the nonlinearity of the dependence on the pdf's, these endpoint quantities are not the equivalent of averaging the full path EAA over trajectories with fixed endpoints. We will explore this distinction further below. 

As was done for endpoint entropy production in \cite{weiss2007fluctuation}, we consider the average over endpoints for an N-d OU system and obtain:
\begin{align}
    \langle \sigma^e \rangle_e &= \mathrm{tr}[\mathbf{C}_t^{-1}(e^{\mathbf{F}t} \bC_0e^{\mathbf{F}^Tt}-e^{2\mathbf{F}t}\bC_0)], \label{eq:sigma_endpoint_average}\\
    \langle \phi^e \rangle_e &= \frac{1}{2}\langle \sigma^e \rangle_e + \frac{1}{2}\ln \det(\bC_0 \bC_t) \nonumber \\  
    &\qquad + N[1 + \ln(2\pi)], \label{eq:phi_endpoint_average} \\
     \langle \alpha^e \rangle_e &= \frac{1}{2}\ln \det(\bC_0 \bC_t) + N[1 + \ln(2\pi)].
     \label{eq:alpha_endpoint_average}
\end{align}
Details can be found in Appendix \ref{appendix:average_quantities}. Note that while $\langle \sigma^e \rangle_e$ is coordinate invariant, the other two quantities are not.

The variations with respect to the climatology process read:
\begin{align}
    \langle \Delta \sigma^e \rangle_e &=  \mathrm{tr}[\mathbf{C}_t^{-1}(e^{\mathbf{F}t} \bC_0e^{\mathbf{F}^Tt}-e^{2\mathbf{F}t}\bC_0)],\label{eq:delta_sigma_endpoint_average}\\
    \langle \Delta \phi^e \rangle_e &= \frac{1}{2} \langle \Delta\sigma^e \rangle_e + \frac{1}{2} \ln \det(\bC_t \bC_0^{-1}), \label{eq:delta_phi_endpoint_average} \\
     \langle \Delta \alpha^e \rangle_e &= \langle \Delta \phi^e \rangle_e - \frac{1}{2}\langle \Delta \sigma^e \rangle_e,\\
     &=\frac{1}{2}\ln \det(\bC_t \bC_0^{-1})\label{eq:delta_alpha_endpoint_average}.
\end{align}
These quantities are now coordinate invariant and thus depend only on the dynamical parameters in $(\bI,\bLambda_\bD,\bL)$ coordinates. 

Since $\bC_0=\bI$ in $(\bI,\bLambda_\bD,\bL)$ coordinates, lengths have been nondimensionalized by the covariance. We now refocus on $N=2$. The dynamical parameters $(\bar{d},\Delta d, \ell)$ all have units of time$^{-1}$. If one has already fixed the time unit by a discrete observation timestep in the LIM algorithm, the parameters can no longer be nondimensionalized. However, one could instead choose any of $\bar{d}$, $\Delta{d}$, or $\ell$ to define a timescale, leaving two dynamical parameters. Below we will find that overall diffusive decay is governed by $\bar{d}$, motivating us to choose a timescale $\tau = 1/\bar d$ in our analysis of OU dynamics. The observational timestep is then $\Delta \tilde{t} = \Delta t/\tau$. If we wish to observe more than one point on individual fluctuations before they diffusively decay, one needs $\Delta \tilde{t} < 1$. The nondimensional dynamical parameters are then $\lambda = \ell/\bar d$ and $\delta = \Delta d/\bar d$. In later sections where we discuss the MJO application, we will instead use the original three parameters ($\bar d$, $\Delta d$, $\ell$) and keep $\Delta t = 1$ day, since MJO parameter evolution over the twentieth century is of direct physical interest.

For $N = 2$, the quantities now depend only on the two parameters $\lambda$, $\delta$, and their combination in $\tilde{\omega} = \omega/\bar d =\sqrt{\lambda^2-\delta^2}$:
\begin{widetext}
    
\begin{align}
    \langle \sigma^e \rangle_e (\tilde t)&= \frac{2\lambda^2e^{-2 \tilde{t}}[\cos(2\tilde \omega \tilde{t})-1][1 - e^{-2 \tilde{t}}]}{e^{-2 \tilde{t}}[2\lambda^2-2\delta^2\cos(2\tilde \omega \tilde{t})-\tilde \omega^2e^{-2 \tilde{t}}]-\tilde \omega^2},\label{eq:reduced_CDL_sigma} \\
    \langle \Delta \phi^e \rangle_e (\tilde t)&= \frac{1}{2} \langle \sigma^e \rangle +\frac{1}{2}\ln\left[1-\frac{2e^{-2\tilde{t}}}{\tilde \omega^2}\left(\lambda^2-\delta^2\cos(2\tilde \omega \tilde{t})\right)+e^{-4\tilde{t}}\right], \label{eq:reduced_CDL_phi} \\
    \langle \Delta \alpha^e \rangle_e (\tilde t)&= \frac{1}{2} \ln\left[1-\frac{2e^{-2\tilde{t}}}{\tilde \omega^2}\left(\lambda^2-\delta^2\cos(2\tilde \omega \tilde{t})\right)+e^{-4\tilde{t}}\right]. 
    \label{eq:reduced_CDL_alpha}
\end{align}

\end{widetext}
\noindent where we defined $\tilde{t} = \bar d \,t = t/\tau$. Details are in Appendix \ref{appendix:red_cdl}. Real $\tilde \omega$ corresponds to complex eigenvalues of $\bF$ and oscillatory thermodynamic quantities.

It is also interesting to consider the full path entropy production rate. One can show \cite{weiss2007fluctuation} that the average full-path entropy production rate is equal to the zero-time growth rate of the endpoint-averaged endpoint entropy production. In the $(\bC_0,\bD,\bL)$ representation, the average entropy production rate is
\begin{align}
    \langle \dot \sigma \rangle_\bX &= \left.\frac{d \langle \sigma^e_t \rangle_e}{dt}\right|_{t=0} = \mathrm{\tr}[\mathbf{L}^T\mathbf{D}^{-1}\mathbf{L}\mathbf{C}_0^{-1}], \nonumber \\
    & = \frac{2\bar d}{\bar d^2 - \Delta d^2}\ell^2 = \frac{1}{\tau} \frac{2\lambda^2}{1 - \delta^2},
    \label{eq: ent_prod_rate_3}
\end{align}
where the last 2 equalities hold for $(\bI,\bLambda_\bD,\bL)$ coordinates. In equilibrium, $\ell = \lambda=0$, and $ \langle \dot \sigma \rangle_\bX$ is manifestly zero. The average full-path entropy production grows as either the rotation or diffusion anisotropy grows, and decays as the mean diffusive time grows. 

In this section, we have focused on entropy production, dynamical activity, and action defined directly from the endpoint pdfs. These endpoint-based definitions are motivated by the fact that observations provide discrete trajectories, while the full continuous paths are generally unmeasurable. Conceptually, however, one can instead consider the average of the path-dependent thermodynamic quantities over all trajectories connecting the {\it same} endpoints. In the next section, we establish the connection between these two constructions using information theory.

\subsection{Information Theory interpretation}
\label{subsec:info_theory}

Information theory provides a rigorous framework for quantifying uncertainty in complex systems, and has found widespread application in climate science. Its tools, such as Shannon entropy, mutual information, and relative entropy, offer predictability measures and have been used to assess the limits of forecast skill, detect causal relationships, and characterize interactions within and between components of the climate system \cite{delsole2004predictability, delsole2005predictability, roulston2002evaluating}. These metrics are particularly valuable in capturing nonlinear and non-Gaussian dependencies, where traditional statistical methods often fail. Information theory quantifies properties of arbitrary pdfs. 

Here we apply information theory to pdfs motivated by predictability and stochastic thermodynamics. We summarize results whose derivations can be found in Appendices \ref{appendix:info_theory} and \ref{appendix:inequalities}. In information theoretic terms, the averages of Eqs.~\eqref{eq:coarse_grained_sigma} - \eqref{eq:coarse_grained_phi} are  (see Appendix \ref{appendix:info_theory}):
\begin{align}
    \langle \sigma^e \rangle_e 
    &= \mathcal{D}_{KL}(P^e\|\hat P^e), \\
    \langle \phi^e \rangle_e &=  \mathcal{H}[P^e] + \frac{1}{2} \mathcal{D}_{KL}(P^e\|\hat P^e)  \\
    \langle \alpha^e \rangle_e &=  \mathcal{H}[P^e]
\label{eq:info_theory_coarse_graining_eqs}
\end{align}
while the variations with respect to $P^e_{ref}(\bx_0, \bx_1) = P_0^e(\bx_0, \bx_1) = P_0(\bx_0)P_0(\bx_1)$ are:
\begin{align}
    \langle \Delta \sigma^e \rangle_e &= \langle \sigma^e \rangle_e \\
     \langle \Delta \phi^e \rangle_e &= - \mathcal{D}_{KL}(P^e\|P^e_0) + \frac{1}{2} \mathcal{D}_{KL}(P^e\|\hat{P}^e),  \\
    \langle \Delta \alpha^e \rangle_e &= - \mathcal{D}_{KL}(P^e\|P^e_0)
     \label{eq:info_theory_coarse_graining_eqs_1}
\end{align}
where $\mathcal{H}$ is the Shannon entropy and $\mathcal{D}_{KL}$ the Kullback-Leibler divergence, which is a measure of the difference between two distributions. Note that, although the activity $\phi^e$ is time-symmetric at the level of individual trajectories, its ensemble average is not necessarily so because it is computed with respect to the forward distribution, which itself breaks time-reversal symmetry. This is manifested in the non-symmetric nature of the KL divergence. We refer the reader to Appendix \ref{appendix:info_theory} for the case of a generic $P^e_{ref}$. These relations make explicit the coordinate-invariant properties of $\langle \Delta \sigma^e \rangle_e$, $\langle \Delta \phi^e \rangle_e$ and $\langle \Delta \alpha^e \rangle_e$, compared to $\langle  \phi^e \rangle_e$ and $\langle  \alpha^e \rangle_e$. 

We are now in a position to compare the endpoint quantities, built from $P^e$, with the fixed-endpoint average of the full-path quantities. One can obtain the following equalities \cite{weiss2009nonequilibrium} (see Appendic \ref{appendix:inequalities})
\begin{align}
    \langle e^{-\sigma}\rangle_{\bX^e} &= e^{-\sigma^e}, \nonumber \\
    \langle e^{\Delta \alpha} \rangle_{\bX^e} &= e^{\Delta \alpha^e}.
    \label{eq:exp_sigma_action_total_vs_endpoints}
\end{align}
The relationships between the EAA quantities themselves, rather than their exponentials, are obtained from the decomposition of the path probabilities in Eqs. \eqref{eq:decomposition_forward_path_prob} and \eqref{eq:decomposition_backward_path_prob}:
\begin{align}
\langle \sigma \rangle_{\bX^e} &= \sigma^e + \mathcal{D}_{KL}(P^{\bX^e}\| P^{\hat{\bX}^e}), \nonumber \\
 \langle \phi \rangle_{\bX^e} &= \phi^e + \mathcal{H}(P^{\bX^e}) 
    +\frac{1}{2}\mathcal{D}_{KL}(P^{\bX^e}\| P^{\hat{\bX}^e}), \nonumber\\
    \langle \Delta \alpha \rangle_{\bX^e} &= \Delta \alpha^e - \mathcal{D}_{KL}(P^{\bX^e}\| P_{ref}^{\bX^e}).
\label{eq:all_total_vs_endpoints}
\end{align}
Since the KL divergence is positive or zero, or, alternatively, applying Jensen's inequality to Eq.(\ref{eq:exp_sigma_action_total_vs_endpoints}), we deduce
\begin{align}
\sigma^e &\leq \langle \sigma \rangle_{\bX^e},\\
\Delta \alpha^e &\geq \langle \Delta \alpha \rangle_{\bX^e}
\label{eq:all_total_vs_endpoints_1}
\end{align}
This is consistent with the result in \cite{esposito2012stochastic}, which showed that coarse-graining over microscopic states generally reduces entropy production, such that the coarse-grained entropy production provides a lower bound on the total entropy production. There is no similar inequality for the activity since the Shannon entropy is not sign-definite. See Appendix \ref{appendix:inequalities} for more details.

This information-theoretic formulation allows us to establish clear inequalities relating full path-dependent and endpoint quantities, while providing a general framework that is independent of the specific system under consideration.

\section{Potential Predictability}
\label{sec:predictability} 

A fundamental question in forecasting is how much uncertainty is generated by the dynamics itself, versus uncertainty due to model and observational error. Deterministic chaos and stochastic fluctuations continuously amplify small perturbations, causing ensembles to spread and information about the initial state to be lost \cite{lorenz1963deterministic,lorenz1975climatic}. Potential predictability addresses this question in an idealized, perfect-model/perfect-observation framework, where the limits to prediction arise exclusively from the underlying dynamics.

Measures of climate predictability quantify the difference of two probability distributions: the \textit{steady-state} distribution, also called the \textit{reference}, or \textit{climatological} distribution, and the \textit{forecast} distribution \cite{delsole2004predictability, delsole2005predictability}. An event is considered to be unpredictable if the climatological distribution is the only basis for prediction. If a forecast provides additional information beyond climatology, the system is then said to be predictable.

A widely used measure of potential predictability is the Anomaly Correlation Coefficient (ACC) \cite{MurphyEpstein1989,du2024increase}. In the perfect-model/perfect-observation framework, the ACC is the correlation between each ensemble member and the corresponding ensemble mean. Consider an ensemble forecast initialized at time $t_i$ at the perfectly observed state $\bx(t_i)$. The initial distribution is then a $\delta$-function at $\bx(t_i)$. Each ensemble member $n$ follows a trajectory $\bx^{n}(t;t_i)$. After a forecast lead time $t_f$, the ensemble mean forecast, $\bar \bx$, is verified against observations of the ``truth''. In the perfect-model framework, each ensemble member is taken alternatively as "truth". A general formulation of the ACC is then (see \cite{du2024increase}):
\begin{equation}
    ACC= \frac{\mathbb{E}[\bx^n_f \cdot \bar \bx_f]}{\sqrt{\mathbb{E}[\bx^n_f\cdot \bx^n_f] 
    \mathbb{E}[\bar \bx_f\cdot\bar \bx_f]}},
    \label{eq:ACCdef}
\end{equation}
where $\bx^n_f$ is the forecast state of the $n^{th}$ ensemble member, $\bar \bx_f$ is the ensemble mean forecast from a single initialization, and $\mathbb{E}[ \, \cdot \, ]$ is an average over both the forecast ensemble for a single initialization and over an ensemble of initial conditions chosen from the steady-state. For an OU ensemble forecast system, the ensemble trajectories are given by Eq.~\eqref{eq:ou_solution}, where each ensemble member is perturbed by independent noise and the ensemble mean is $\bar \bx(t;t_i) = e^{\bF (t- t_i)}\bx(t_i)$. The perfect-model/perfect-observation ACC with forecast lead time $t$ for OU dynamics is then (see \ref{appendix:acc_langevin})
\begin{equation}
    ACC(t) = \sqrt{\frac{\tr[\mathbf{C}_{0} - \mathbf{C}_{t}]}{\tr[\mathbf{C}_{0}]}}.
    \label{eq:acc_langevin}
\end{equation}
At $t=0$, perfect observations give a $\delta$-fn ensemble, $\bC_{t = 0} = 0$, and $ACC(0) = 1$. As $t \to \infty$, $\bC_{t} \to \bC_0$ and $ACC(0) \to 0$. The predictability time is often taken to be the time when the ACC falls below 0.6. Note that the ACC is only coordinate invariant under orthogonal coordinate transformation.
Nonetheless, it is interesting to consider $(\bI,\bLambda_{\bD}, \bL)$ coordinates and scale time by $\bar d$, obtaining
\begin{equation}
    ACC(\tilde{t}) = e^{-\tilde{t}}
    \sqrt{\frac{\lambda^2 - \delta^2\cos(2 \tilde{\omega}\tilde{t})}{\lambda^2 - \delta^2}}.
    \label{eq:acc_CDL}
\end{equation}
where $\tilde \omega^2 = \lambda^2 - \delta^2$ again. See Appendix \ref{appendix:acc_langevin} for details.

In $(\bI,\bLambda_{\bD}, \bL)$ coordinates, a purely rotational deterministic system with $\bD = 0$ has $ACC = 1$ for all times. If, instead, we consider fixed $\delta$ and vary $\lambda$, then the slowest ACC decay occurs when $\lambda = 0$, i.e., in thermodynamic equilibrium.

Both the ACC, Eq. \eqref{eq:acc_langevin}, and the average endpoint action, Eq. \eqref{eq:delta_alpha_endpoint_average}, with a climatological reference, depend on the OU dynamics through $\bC_0$ and $\bC_t$. The action is thus suggested as another measure of predictability. In the case of isotropic diffusion in $(\bI,\bLambda_{\bD}, \bL)$ coordinates, the distinction collapses and
\begin{align}
    (ACC)^2(t) &= 1 - e^{\langle \Delta \alpha^e_t \rangle_e} \quad \hbox{if } (\bC_0 = \bI, \, \Delta d = 0).
    \label{eq:alpha_pred}
\end{align} 
This motivates defining an action-based measure of predictability, 
\begin{equation}
    \rho = 1-e^{\langle \Delta \alpha^e \rangle_e},
\end{equation}
which,  like the ACC, is bounded between 0 and 1, but has the advantage of being coordinate invariant. We can decompose the action-predictability into
\begin{equation}
    \rho = 1 - e^{\langle \Delta \phi^e \rangle -\frac{1}{2}\langle \Delta \sigma^e \rangle}.
\end{equation}
This highlights that predictability, irreversibility, and dynamical activity capture distinct aspects of the dynamics.

Below we will also consider additional measures of potential predictability: the Root Mean Square Error (RMSE), phase error, and amplitude error (refer to Appendix \ref{appendix:skill_scores} for details). For a Langevin system, the RMSE is $\sqrt{\mathrm{tr}[\bC_t]}$, and, by normalizing by $\sqrt{\mathrm{tr}[\bC_0]}$, we get that $ACC = \sqrt{1 - RMSE^2}$.

\section{Rotational and diffusive regimes}
\label{sec:rot_diff_regimes}

Exploration of the parameter space reveals distinct dynamical regimes whose characteristics depend on the values of $\lambda$ and $\delta$. The sign of $\lambda$ determines the direction of rotation, but does not change the regime of the system. 

The character of the eigenvalues of the drift, $\mu_\bF$, divides the dynamics into two regimes: the diffusive regime where $\lambda^2 < \delta ^2$, $\tilde \omega$ is imaginary, and the eigenvalues $\mu_\bF$ are real; and the rotational regime where $\lambda^2 > \delta^2$, $\tilde\omega$ is real, and $\mu_\bF$ are complex. 

The Green function, $\bG(\tilde{t})= \exp(\bF\tilde{t})$, is
\begin{widetext}
    
\begin{equation}
    \bG(\tilde{t}) = e^{-\tilde{t}}
    \begin{bmatrix}
     \cos (\tilde{\omega} \tilde{t} )-\frac{\delta}{\tilde{\omega}} \sin(\tilde{\omega} \tilde{t}) &
     -\frac{\lambda}{\tilde{\omega}} \sin(\tilde{\omega} \tilde{t}) \\
     \frac{\lambda}{\tilde{\omega}} \sin (\tilde{\omega} \tilde{t}) & \cos(\tilde{\omega} \tilde{t})+\frac{\delta}{\tilde{\omega}}
        \sin (\tilde{\omega} \tilde{t})
    \end{bmatrix}.
    \label{eq:Gddl}
\end{equation}

\end{widetext}
In the rotational regime with $\bar{d}$ temporal scaling, the overall exponential decay rate of the Green function is 1,  and the matrix elements oscillate as $\tilde{\omega} \tilde{t}$. In the diffusive regime, the trigonometric functions in $\bG$ become hyperbolic trig functions, which contribute with an additional exponential behavior. The exponential growing piece of $\cosh$ and $\sinh$ grow as $\exp(\tilde{\omega} \tilde{t})$. Since by definition $\delta < 1$, this growth is slower than the leading exponential decay, and the impact of $\tilde{\omega}$ in the diffusive regime is to slow the diffusive decay of the Green function.

Note that the diffusive regime can still have $\lambda \ne 0$ and is thus a nonequilibrium regime with mean phase space rotation, despite the fact that $\bG$ has no explicit rotation. In the rotational regime, $\bG$ has explicit rotation with frequency $\tilde{\omega} \le |\lambda|$. In both regimes, the probability current velocity $\bv =  -\bL \bC_0^{-1}\bx$ and is independent of the diffusion.

The special case of isotropic diffusion, where for general coordinates $\bD \propto \, \bC_0$, results in significant simplifications. Note that while both the choice of representation and the coordinate transformation to $(\bI,\bLambda_\bD,\bL)$ coordinates is completely general, demanding $\bD \propto \, \bC_0$ is a significant constraint on the dynamics. In $(\bI,\bLambda_\bD,\bL)$ coordinates, this becomes $\bD = \bar{d}\,\bI$ and $\Delta d = 0$. In this  regime
\begin{align}
    \bF &= \begin{pmatrix}
      -\bar{d} & -\ell\\
      \ell & -\bar{d}
    \end{pmatrix},\\
    \bG &= e^{-\bar{d} t} \bR(\ell t) = e^{-\tilde{t}} \bR(\lambda \tilde{t}) 
\end{align}
where $\bR$ is the usual 2d rotation matrix
\begin{align}
    \bR(\theta) &= \begin{pmatrix}
      \cos(\theta) & -\sin(\theta)\\
      \sin(\theta) & \cos(\theta)
    \end{pmatrix}.
 \end{align}
Thus, with isotropic diffusion, the exponential decay rate is 1, as the system is always in the rotational regime, and $\bG$ induces pure phase space rotation with angular frequency $\lambda$. 

\begin{figure}
\centering
\includegraphics[width=0.8\linewidth]{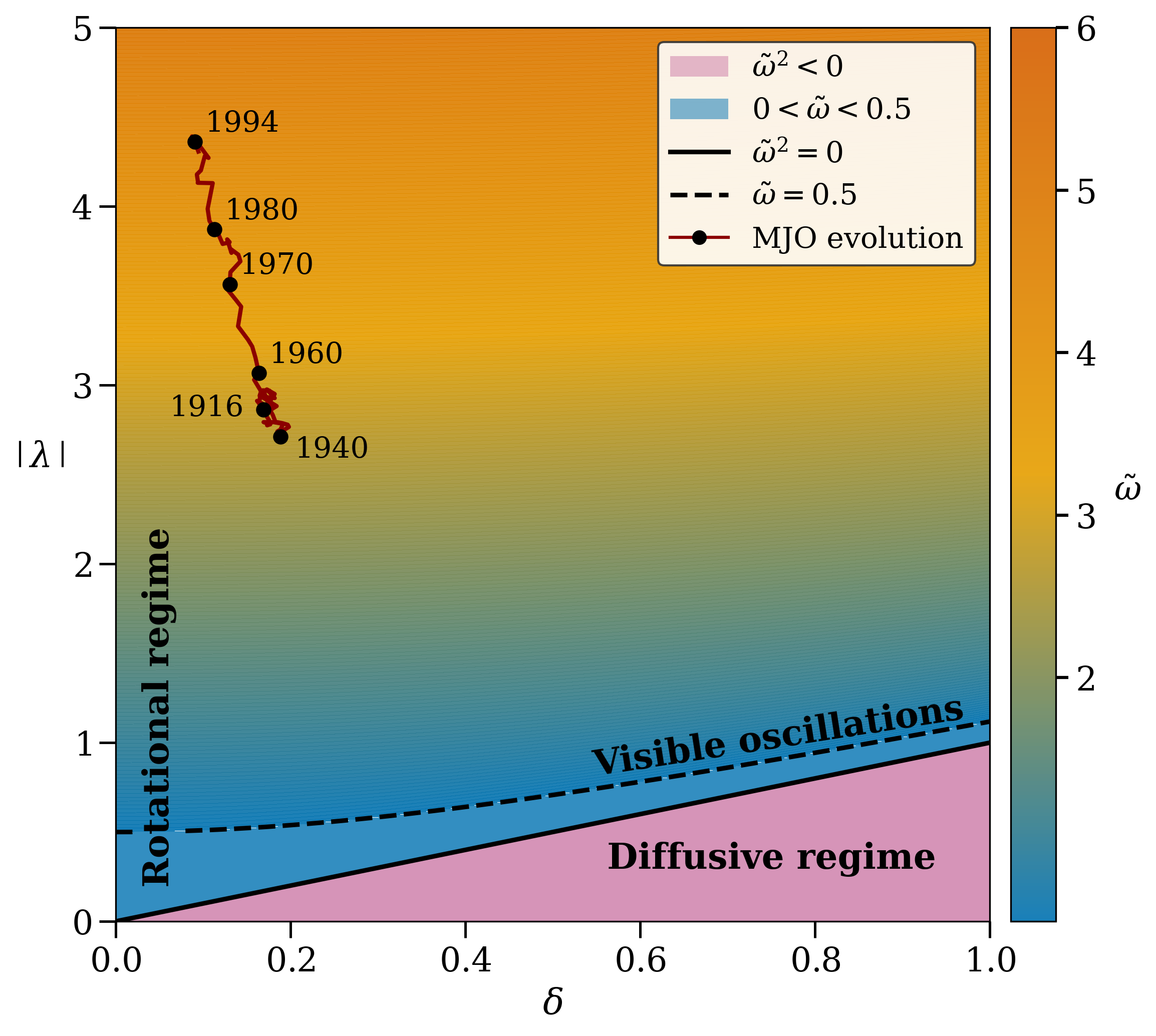}
\caption{\textbf{Parameter space diagram.} Regimes are classified in terms of $|\lambda|$ and $\delta$. The colorbar shows $\tilde{\omega}$ in the rotational regime ($\tilde{\omega} \in \mathbb{R}$), corresponding to oscillatory solutions. The solid curve denotes $\tilde{\omega}^2 = 0$, separating rotational and diffusive regimes, while the dashed curve marks the onset of visually discernible oscillations ($\tilde{\omega} = 0.5$). The temporal evolution of MJO parameters over the 20th century is also shown.
\label{fig:regimes_scheme}}
\end{figure}

All 2d OU steady states can be organized into this 2d $(\delta,\lambda)$ dynamical parameter space, summarized in Fig.~\ref{fig:regimes_scheme}. Three lines in this space have special meaning. Equilibrium steady states, $\lambda = 0$, lie on the horizontal axis of the diagram, NESSs with isotropic diffusion lie on the vertical axis, and the line $\lambda^2 = \delta^2$ defines the boundary between the diffusive and rotational regimes. 
    
From Eqs.~\eqref{eq:reduced_CDL_sigma} - \eqref{eq:reduced_CDL_alpha} one sees that the averaged endpoint path functions have terms like $\exp(-2 \tilde{t})\cos(2\tilde{\omega} \tilde{t})$. In the region of the rotational regime where $\tilde{\omega} \ll 1$, the exponential decay largely reduces the functions to zero before oscillations become visible. For sufficiently large $\tilde{\omega}$, however, visible oscillations are apparent. Just how large oscillations must be to be visible is somewhat subjective, so this crossover is a fuzzy region, but $\tilde{\omega} \sim 0.5$ is a reasonable boundary. In addition, since EAA are quadratic in $\bG$, both timescales, $\tau$ and $1/\omega$, are reduced by a factor of 2 as compared to those of the underlying dynamics, resulting in $\langle \sigma^e \rangle_e = 0$ at half a cycle, i.e., forward and backward evolutions become statistically indistinguishable there.

\begin{figure}
\centering
\includegraphics[width=\columnwidth]{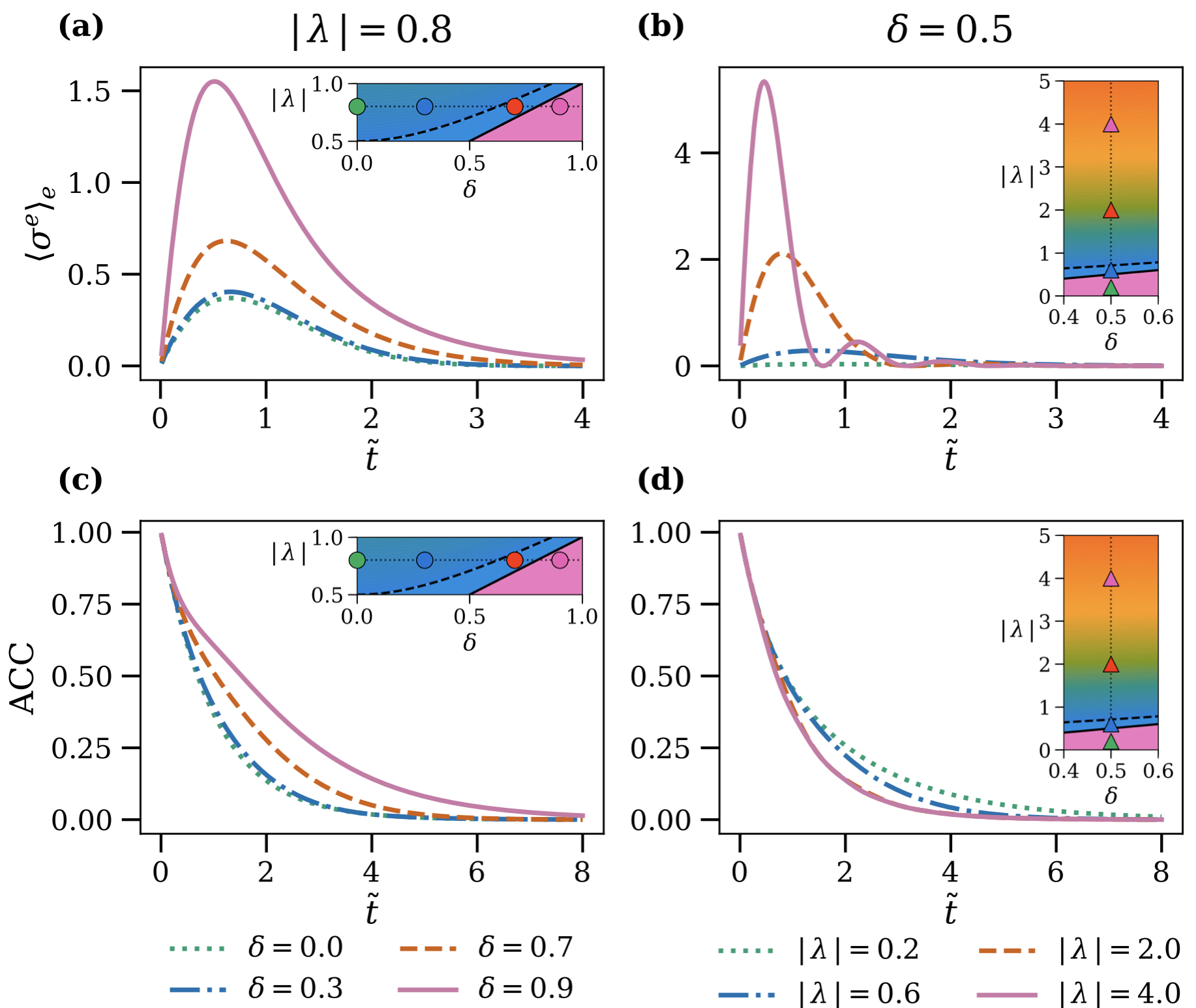}
\caption{\textbf{Dependence of entropy production and ACC on $\boldsymbol{\lambda}$ and $\boldsymbol{\delta}$.} Panels show the time evolution of $\langle \sigma_t^e \rangle_e$ (\textbf{a}--\textbf{b}) and the ACC (\textbf{c}--\textbf{d}). The first column varies $\delta$ at fixed $\lambda$, and the second varies $\lambda$ at fixed $\delta$. Time is scaled as $\tilde{t} = t/\tau$. Insets show the corresponding locations in parameter space across all regimes. \label{fig:regimes_delta_lambda}}
\end{figure}

In the diffusive regime, the ACC decays more slowly as well due to the countervailing effect of the growing exponential $e^{2f\tilde{t}}$ (from the cosine functions with imaginary arguments), where $f = i\tilde{\omega} < 1$ is real and positive. In this case, a system is more predictable the larger $f$ is, that is, the larger $\delta$ is compared to $\lambda$. In the rotational regime, in contrast, the ACC's decay rate is approximately 1 regardless of the values of $\delta$ and $\lambda$. The square root in \ref{eq:acc_CDL} oscillates between 1 and $\sqrt{\frac{\lambda^2 + \delta^2}{\lambda^2 - \delta^2}}$, and, thus, it becomes more influential when $\delta \simeq \lambda$, i.e. $\tilde{\omega} \simeq 0$. Consequently, for highly rotational systems, $\lambda \gg \delta$, the ACC is approximately decaying simply as $e^{-\tilde{t}}$. In the isotropic case, the equality is exact. It is worth noting that $\langle \Delta \alpha^e \rangle$ exhibits a similar insensitivity to changes in $\lambda$ within the rotational regime.

Fig. \ref{fig:regimes_delta_lambda} describes the decay over time of $\langle \sigma^e \rangle_e$ (\textbf{a},  \textbf{b}) and ACC (\textbf{c}, \textbf{d}) for a fixed value of $\lambda$ and varying $\delta$, and viceversa, as one crosses parameter space. The peak value of the average endpoint entropy production increases with $\delta$, whereas increasing $\lambda$ results in more frequent oscillations of greater amplitude. The ACC's decay, instead, decreases with $\delta$ and increases with $\lambda$ in the diffusive regime, but it remains approximately the same in the rotational regime. Thus, in the rotational regime, a system can experience a substantial increase in irreversibility while preserving the same degree of predictability.

\section{Madden-Julian Climate Oscillation}
\label{sec:MJO}

The stochastic thermodynamic framework developed in Sections II–V is fully general and applies to any 2d OU process. The framework can, in principle, be applied to any climate oscillation admitting a low-dimensional phase-space representation. We choose the Madden-Julian Oscillation (MJO) as our test case primarily because its intraseasonal timescale yields far more independent cycles than slower oscillations such as the El-Ni\~no/Southern Oscillation, providing more robust parameter estimates across 30-year sliding windows. Here we compare the OU dynamics to a modern ensemble forecast system in terms of the perfect-model potential predictability of the two models, as well as their actual skill in forecasting the state of the MJO.

\subsection{Datasets}
\label{subsec:dataset}

A Linear Inverse Model is built from a dataset of a set of indices chosen to provide a low-dimensional representation of the oscillation of interest. For the Madden–Julian Oscillation, the 2d Real-time Multivariate MJO Index (RMMI) is commonly used, as it captures the primary eastward-propagating convective patterns along the tropics. The RMMI are based on the first two principal components of the empirical orthogonal functions (EOFs) of filtered outgoing long-wave radiation and upper and lower level zonal winds \cite{WheelerHendon2004}. The length of the RMMI state vector measures the oscillation amplitude, its angle indicates its phase, and the rotation of the state vector in phase space directly mirrors the physical eastward propagation of the convective system along the tropics. 

\begin{figure}
    \centering
    \includegraphics[width=0.7\linewidth]{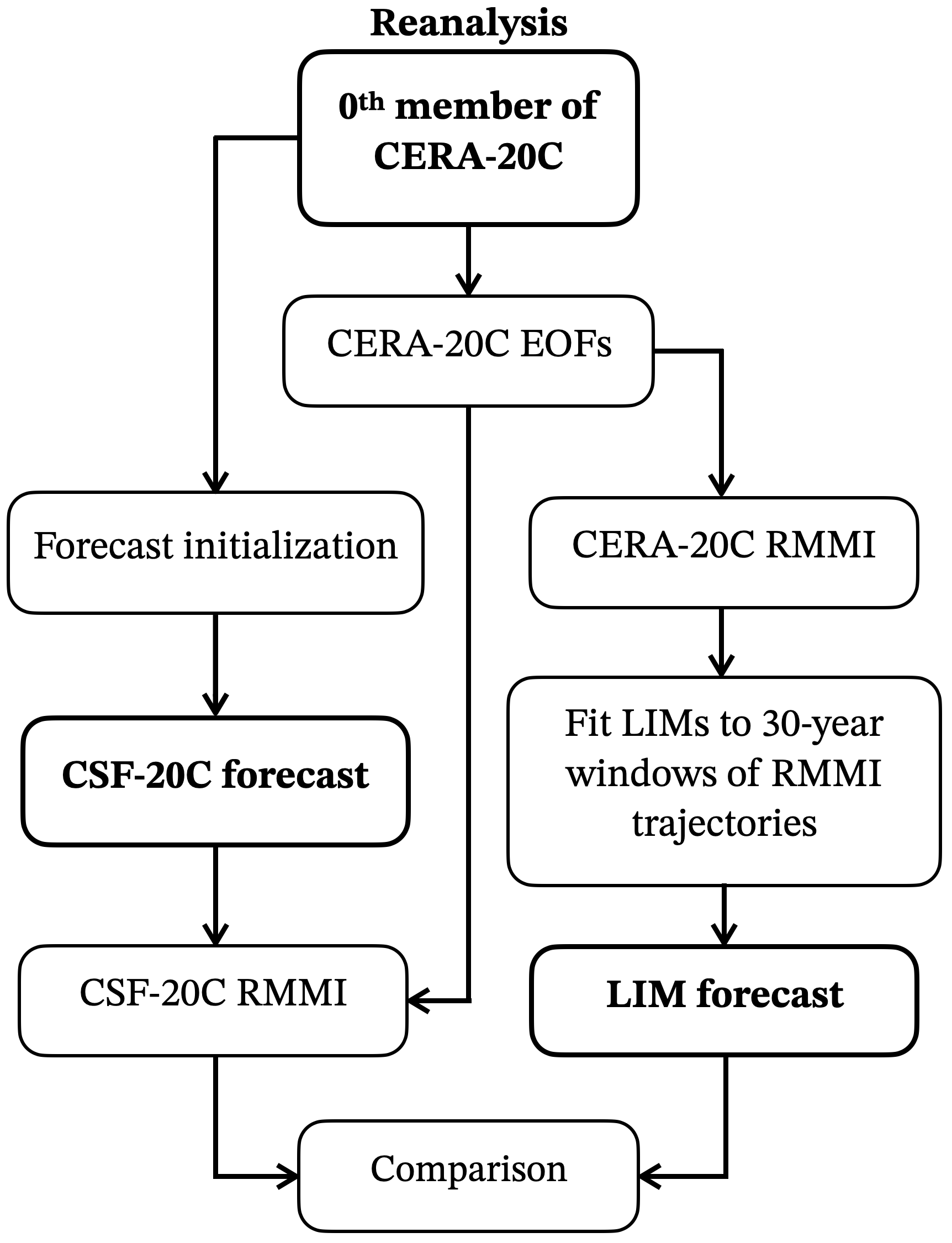}
    \caption{\textbf{Scheme of the comparison between the LIM, the CSF-20C, and the CERA-20C}. The $0^{\mathrm{th}}$ member of CERA-20C reanalysis is used both as the initialization for CSF-20C and, after extracting the RMMI timeseries, as a training set for the LIM. In the case of CSF-20C, the RMMI timeseries is computed from the model output. All RMMI timeseries are built from the EOFs of CERA-20C. Bold texts refer to the forecasts and reanalysis. }
    \label{fig:scheme_datasets}
\end{figure}

To construct a LIM of the MJO, we use the RMMI time series spanning the $20^{\mathrm{th}}$ century. These indices are derived from the ECMWF Coupled Reanalysis of the $20^{\mathrm{th}}$ Century (CERA-20C) \cite{laloyaux2018cera}, a ten-member ensemble covering the 1901--2010 period with daily output. Climate reanalyses combine observations with a dynamical model through data assimilation to provide a physically consistent estimate of the past state of the climate system. The RMMI used here is taken directly from Du et al. (2024) \cite{du2024increase}. They used a fixed EOF basis derived from $0^{\mathrm{th}}$ ensemble member over the period 1979–2001. Because this basis is held fixed throughout the century, the phase-space representation itself does not drift over time, ensuring that any trend cannot be attributed to a shifting coordinate system. Details of the RMMI computation are provided in Du et al. (2024) \cite{du2024increase}. 

A LIM is constructed independently for each CERA-20C ensemble member using a 30-year sliding window. The resulting model parameters are used to generate an ensemble of forecasts by numerical integration of Eq. \eqref{eq:langevin} with the Euler-Maruyama method.

\begin{figure}
    \centering
    \includegraphics[width=0.8\columnwidth]{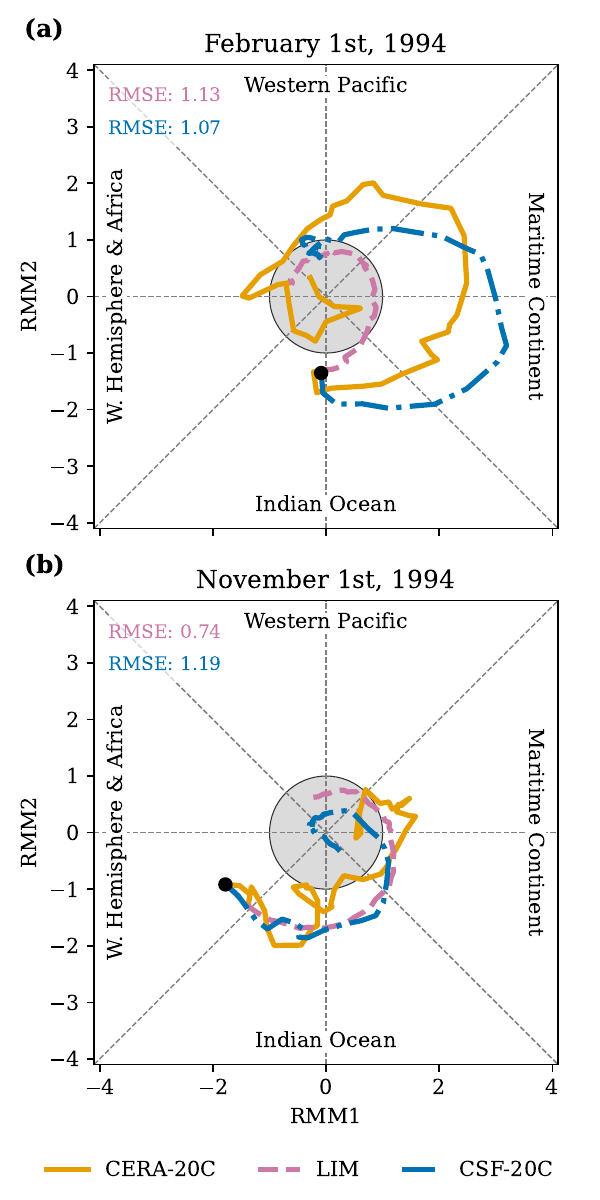}
    \caption{\textbf{MJO phase space trajectories}. 46-day RMMI trajectories from CERA-20C, from the LIM forecast built on the $0^{\mathrm{th}}$ member of CERA-20C ensemble and CSF-20C. Panel \textbf{(a)} shows trajectories initialized on February $1^{\mathrm{st}}$, 1994, while panel \textbf{(b)} those initialized on November $1^{\mathrm{st}}$ of the same year. MJO events are defined by amplitudes greater than one (gray circle). 
    \label{fig:mjo_phase_space}}
\end{figure}

To enable comparison with a more complex forecast system, we also consider RMMI time series derived by Du et al.\cite{du2024increase} from the ECMWF’s Integrated Forecasting System coupled forecast of the 20th Century (CSF-20C) \cite{weisheimer2020seasonal}, projected onto the same EOF basis. These 46-day forecasts are initialized from the $0^{\mathrm{th}}$ member of CERA-20C and include two start dates per year (February and November), with ensemble sizes of 25 and 51 members, respectively, spanning the 1900–2010 period. Seasonal forecasts are model-based ensemble predictions of the average state of the climate system over coming months, initialized from observed conditions and accounting for sources of long-term predictability. We refer the reader to Fig.~\ref{fig:scheme_datasets} for a schematic overview of the datasets and forecast configurations. Two sample MJO phase space trajectories are shown in Fig.~\ref{fig:mjo_phase_space}. In one forecast, the LIM outperforms the CSF-20C in terms of RMSE, while in the other, the  CSF-20C outperforms the LIM.

\subsection{Potential predictability of the MJO}
\label{sub:pot_predictability_mjo}

\begin{figure*}
    \centering
    \includegraphics[width=2.05\columnwidth]{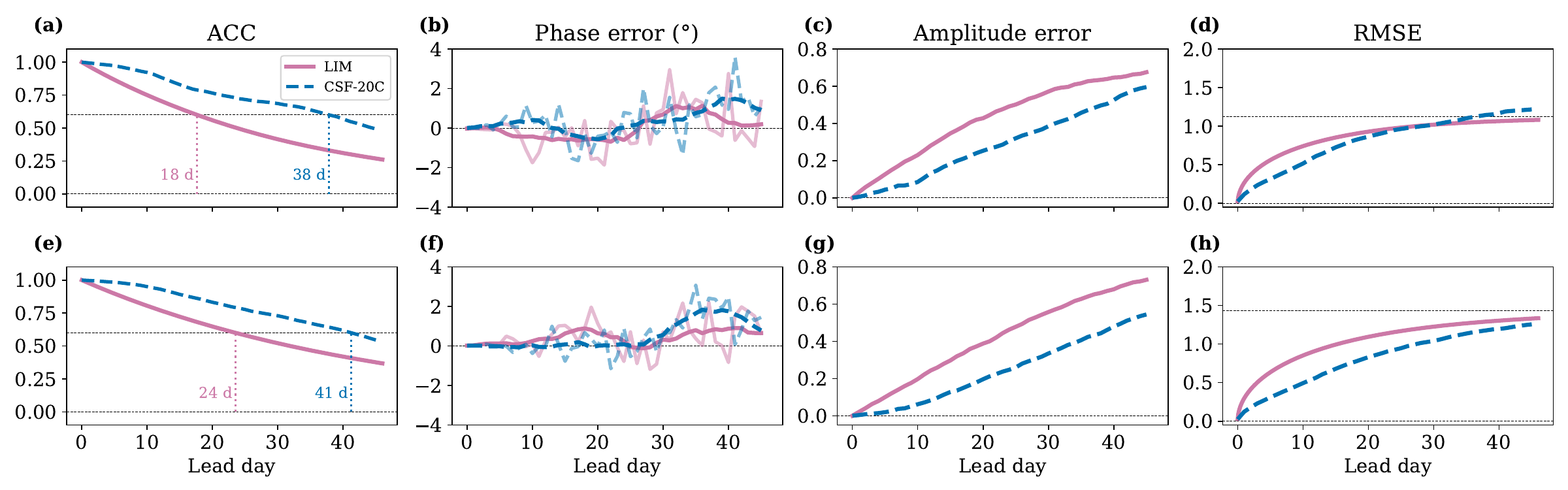}
    \caption{\textbf{Potential predictability according to the two forecasts}. ACC (\textbf{a},\textbf{e}), phase error (\textbf{b},\textbf{f}), amplitude error (\textbf{c},\textbf{g}), and RMSE (\textbf{d},\textbf{h}) of the ensemble-mean forecast against each ensemble member taken alternatively as truth. Full pink line: LIM forecast from the $0^{\mathrm{th}}$ CERA-20C member; dashed blue line: CSF-20C. In the phase error plot, the full line is the smoothed version of the corresponding dashed line. The phase error is computed only when the starting amplitude of the reanalysis is larger than 1. The lead time after which the correlation score first goes below 0.6 is shown in both plots. Dashed lines in the RMSE plots refer to the maximum value for the LIM obtained as $\sqrt{\mathrm{tr}(\bC_0)}$. The top panels refer to the window centered on 1936, the bottom panels (\textbf{e}-\textbf{h}) to 1994. \label{fig:validation_predictability}}
\end{figure*}

Under the perfect-model assumption, the potential predictability of the MJO within each model can be quantified as the mean deviation of ensemble trajectories from their ensemble mean, which represents the intrinsic, irreducible forecast uncertainty arising solely from internal model variability. We explore predictability through a collection of skill scores: ACC, phase error, amplitude error, and RMSE; see Appendix \ref{appendix:skill_scores} for the precise definitions of these metrics. These metrics are intended as uncertainty measures, since they quantify the spread of the ensemble from its mean. For the LIM forecast, the ACC and RMSE can be computed either from an ensemble of stochastic realizations or directly from the covariance matrices of the system, without the need for explicit numerical integration of trajectories. Figure \ref{fig:validation_predictability} indicates that, while the LIM attains approximately 70–80\% of the CSF-20C's ACC during the first three weeks of lead time, its predictability horizon — defined as the first lead time at which ACC falls below 0.6 — is substantially shorter, at around 20 days compared to approximately 40 days for CSF-20C. In addition, the LIM forecast has larger amplitude error and RMSE, but phase error comparable to CSF-20C. \\

\subsection{Assessment of forecast skill}
\label{sub:forecast_skill_mjo}

\begin{figure*}
    \centering
    \includegraphics[width=2.05\columnwidth]{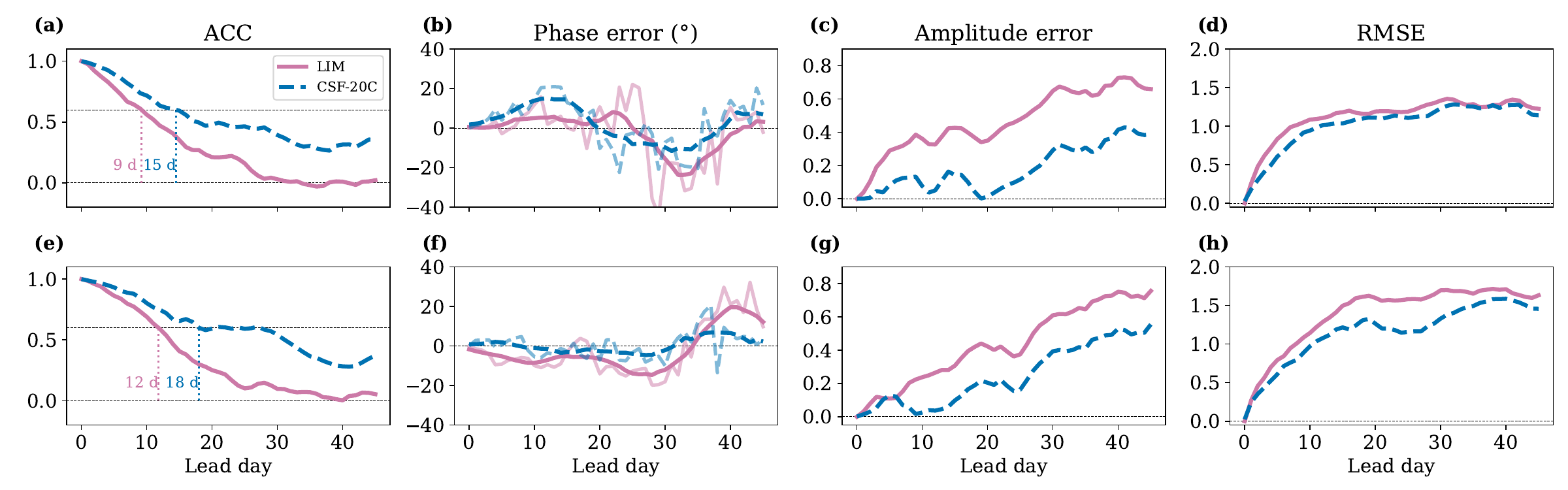}
    \caption{\textbf{Forecast skill scores}. Correlation score computed as ACC (\textbf{a},\textbf{e}), phase error (\textbf{b},\textbf{f}), amplitude error (\textbf{c},\textbf{g}) and RMSE (\textbf{d},\textbf{h}) of the ensemble-mean forecast against the $0^{\mathrm{th}}$ CERA-20C member. Full pink line: LIM forecast from the $0^{\mathrm{th}}$ CERA-20C member; dashed blue line: CSF-20C. In the phase error plot, the dark line is the smoothed version of the corresponding lighter line. The phase error is computed only when the starting amplitude of the reanalysis is larger than 1. The lead time after which the correlation score first goes below 0.6 is shown in both plots. The top panels refer to the 30-year window centered on 1936, the bottom panels (\textbf{e}-\textbf{h}) to 1994. \label{fig:comparison_with_reanalysis}}
\end{figure*}

We compare the performance of the LIM fit to the $0^{\mathrm{th}}$ member of CERA-20C with the CSF-20C by producing an ensemble of LIM trajectories of the same size as the CSF-20C. By comparing the forecast ensemble mean with the reanalysis, we measure the skill of each forecast using a correlation score, an amplitude error, the RMSE, and a phase error, as defined in \cite{rashid2011prediction} and shown also in Appendix \ref{appendix:skill_scores}. Here, the correlation score corresponds to the ACC between the forecast ensemble mean and the observations. We average over both February and November initializations and over each year of the chosen 30-year window. A perfect forecast model would have an ACC equal to one, and phase and amplitude errors together with the RMSE identically zero. Fig. \ref{fig:comparison_with_reanalysis} shows that, although the LIM exhibits lower correlation skill and larger amplitude errors and RMSE, it can match or even outperform CSF-20C in terms of phase. This is consistent with the fact that numerical prediction models still struggle to simulate the phase speed of the MJO accurately, showing a systematically slower phase speed than observed \cite{lim2018mjo}. In particular, panel \textbf{(b)} indicates that the LIM phase error remains smaller than that of CSF-20C up to a lead time of approximately 20 days. Note that the RMSEs and the amplitude errors shown in Fig. \ref{fig:comparison_with_reanalysis} should not be compared across the different 30-year climate periods, as these measures are not coordinate invariant and not normalized. The figure is intended only to compare the two forecast models within each climate period, not to assess how their performance evolves over the century. 

Overall, the LIM generally underperforms the CSF-20C in terms of forecast skill with the exception of phase errors. Nonetheless, OU dynamics captures a substantial fraction of the MJO dynamics and predictability. Remarkably, this is achieved using a linear and highly idealized framework, at a dramatically reduced computational cost, quantifying the extent to which the MJO can be approximated within a simplified three-parameter representation.

\subsection{MJO predictability and EAA}
\label{sec:MJO_results}

Here we explore the entropy production, activity, and action of the MJO. We build an ensemble of LIMs using each member of the CERA-20C dataset and then average over the resulting ensemble of parameters. Then, we transform into the $(\bI,\bLambda_\bD,\bL)$ coordinates, extract $\ell$, $\Delta d$ and $\bar d$, and calculate EAA. In addition, we use a 30-year centered sliding window from 1916 to 1995 in order to measure the evolution of MJO's parameters over the $20^{\mathrm{th}}$ century. 

\begin{figure}
    \centering
    \includegraphics[width=\columnwidth]{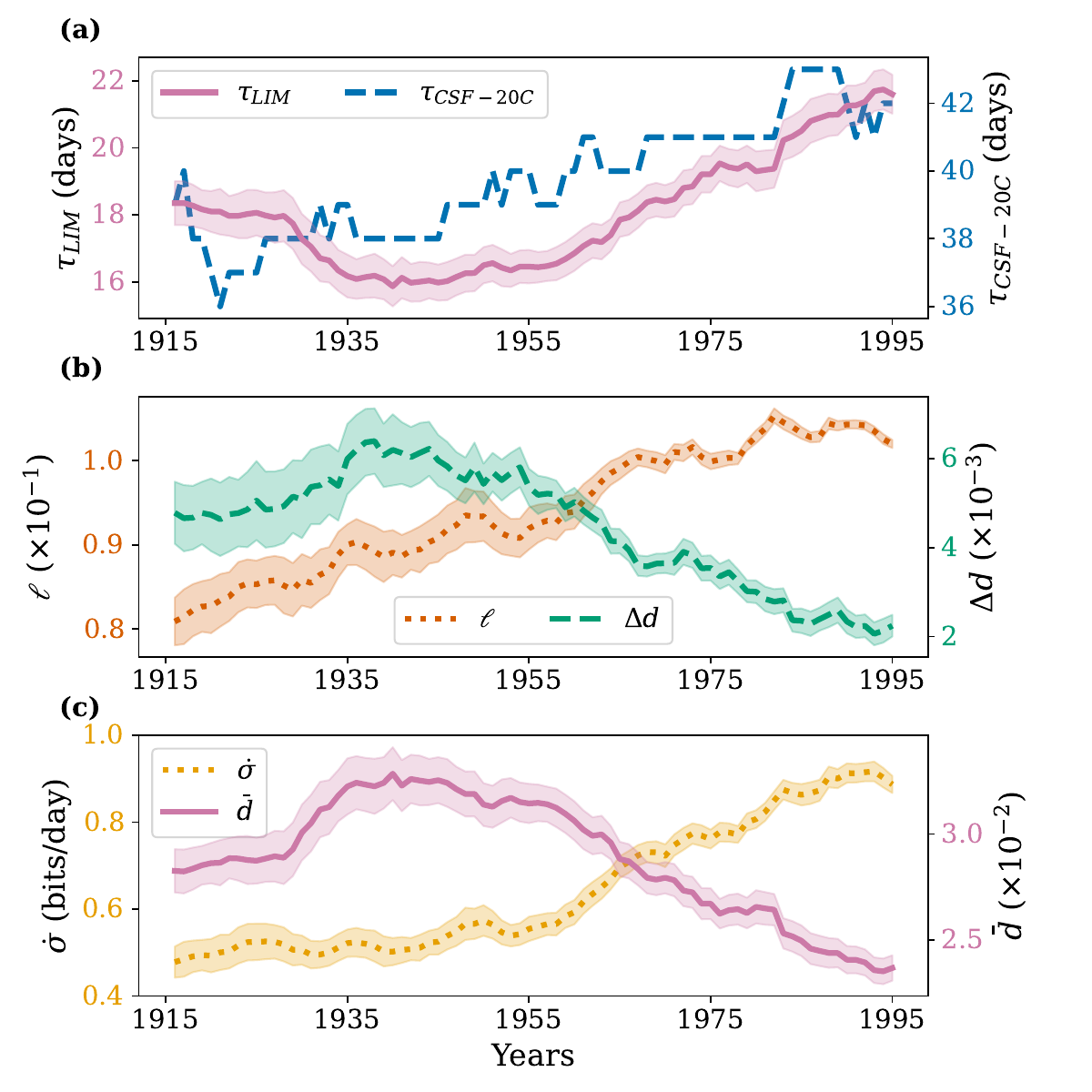}
    \caption {\textbf{Evolution of the MJO over the 20th century}: \textbf{(a)} MJO predictability limit from the CSF-20C, adapted from Du et al (2024) \cite{du2024increase}, and from LIMs of CERA-20C. \textbf{(b)} Corresponding LIM parameters $\ell$ and $\Delta d$. \textbf{(c)} Diffusive rate $\bar d$ and entropy production rate $\dot \sigma$. All quantities use 30-year sliding windows; shaded areas show the spread of the ensemble of LIMs built from the 10 members of the CERA-20C ensemble.\label{fig:MJO_evolution}}
\end{figure}

The top panel of Fig. \ref{fig:MJO_evolution} shows, for each sliding window, the lead times (in days) $\tau_{LIM}$ and $\tau_{CSF-20C}$ at which $ACC_{LIM}$ and $ACC_{CSF}$ first fall below 0.6, respectively. Although the predictability limit captured by the LIM is approximately half that of CSF-20C, spanning about 16 – 23 days compared to 36 – 43 days, respectively, their long-term trends over the twentieth century are similar. Both exhibit higher values at the beginning of the century, reach minima around 1919 (LIM) and 1940 (CSF-20C), and then increase steadily from approximately 1940 onward. This suggests that the evolution of the potential predictability of the MJO is indeed caused to a great extent by linear dynamics. According to both forecast systems, the MJO has become more predictable, as found in Du et al. (2024) with the CSF-20C dataset \cite{du2024increase}. 

In the middle panel, the evolution of $\ell$ and $\Delta d$ over the same period is shown. Probability angular momentum $\ell$ has increased in absolute value over the entire century spanning a range from $0.08$ to $0.105$, meaning that the MJO has become more rotational. On the other hand, $\Delta d$ shows an increase until 1940 and then a steep decline thereafter, going from 0.007 to 0.002, making the MJO more isotropic. Lastly, the bottom panel shows the evolution of the diffusive rate $\bar d$ — which evolves similarly as $\Delta d$ spanning a range from 0.02 to 0.035 — together with the average entropy production rate $\langle \dot \sigma \rangle $, which steadily increases starting from 1940. Thus, the MJO has become more irreversible and less diffusive.

Since $\Delta d$ is respectively one and two orders of magnitude smaller than $\bar d$ and $\ell$ and $\bar d<\ell$, the MJO is in the rotational oscillatory regime, $\omega$ is real, and is approximately isotropic. In this regime, according to Eq. \eqref{eq:acc_CDL} and the results in Section \ref{sec:predictability}, the decay of predictability is almost entirely dependent on $\bar d$. In fact, the diffusive timescale $\tau = 1/(2\bar d)$, with $\bar d$ as in the bottom panel, reproduces almost perfectly the behavior of the LIM predictability curve in the top panel — $e^{-d\tau} = e^{-1/2} \simeq 0.6$. Thus, the increase in predictability of the MJO in its LIM representation is determined nearly exclusively by diffusion, which in this context could represent the effect of unresolved sub-daily (e.g., hourly) processes. In contrast, the entropy production rate depends on all three parameters. It remains approximately constant up to 1940 due to the compensating effects of their simultaneous increase, and subsequently rises as a result of the increase in $\ell$ and the decrease in $\bar d$ and $\Delta d$ (Eq. \ref{eq: ent_prod_rate_3}). 

The century-long trends in the LIM parameters systematically exceed the spread across the 10 CERA-20C ensemble members at every 30-year window, indicating that they are not simply an artifact of internal ensemble variability. This spread itself narrows over the century, consistent with improving observational constraints in more recent decades, which does not undermine the statistical robustness of the trends, but indicates that a residual influence of the evolving observing system on the absolute parameter values and the predictability cannot be fully excluded.

\begin{figure}
    \centering
    \includegraphics[width=0.8\columnwidth]{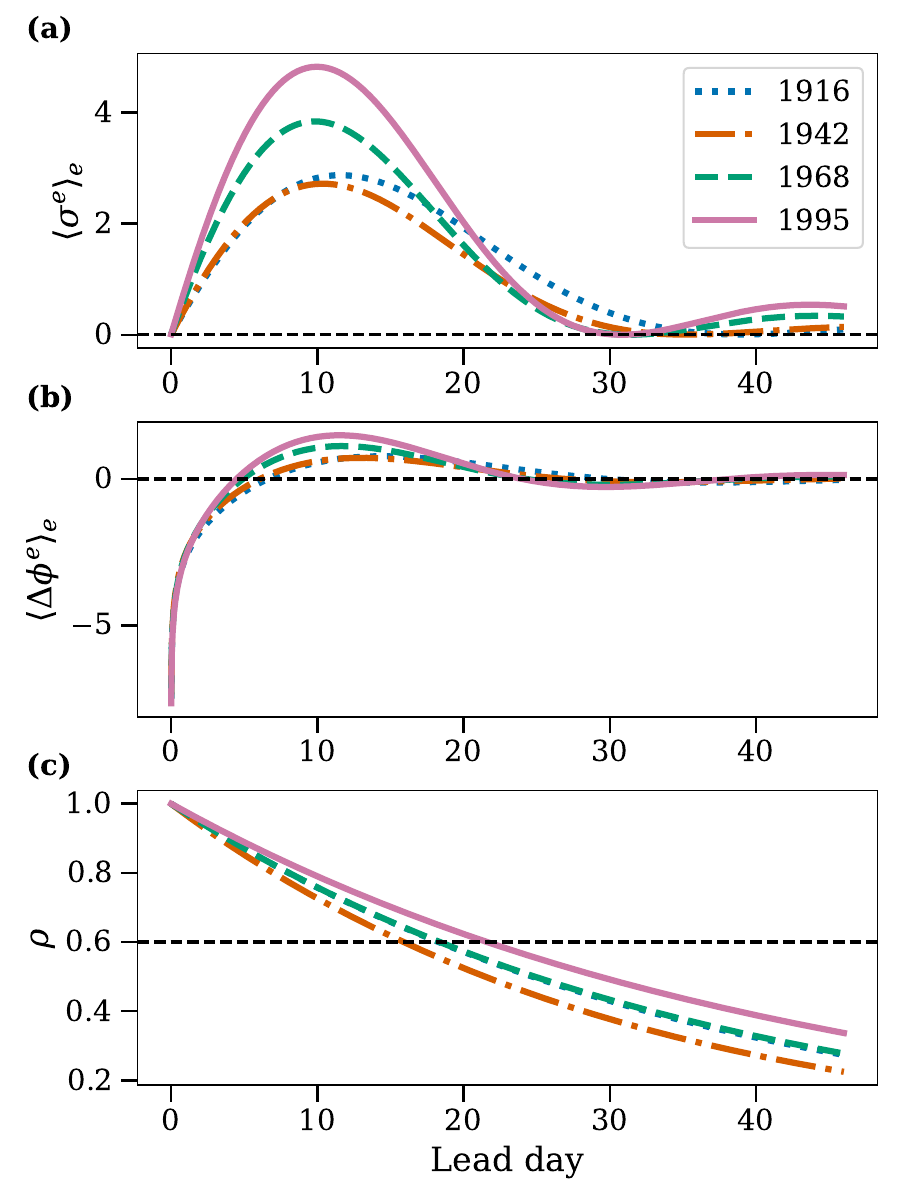}
    \caption{\textbf{EAA of the MJO}: $\langle \sigma^e \rangle_e$ \textbf{(a)}, $\langle \Delta \phi^e \rangle_e$ \textbf{(b)}, and $\rho$ \textbf{(c)} as a function of lead day for 30-year climate windows centered on 1916, 1942, 1968, 1995. The quantities are evaluated from their analytic expressions using OU parameters estimated from CERA-20C through the LIM.\label{fig:thermo_pred_evolution}}
\end{figure}

In a similar vein, Fig. \ref{fig:thermo_pred_evolution} illustrates the evolution of EAA over the 20th century as a function of lead time: $\langle \sigma^e \rangle_e$ and $\langle \Delta \phi^e \rangle_e$ oscillate with increasing amplitude, mainly due to decreasing $\bar d$ that defines the exponential decay, and a decreasing period caused by increasing $\ell$, over the century. Thus, the MJO has become more active as well as more irreversible over the last century. Moreover, due to the approximately isotropic character of the MJO, the two predictability measures, ACC and $\rho$, are essentially the same, as shown in Section \ref{sec:predictability}, with differences of the order of $10^{-3}$. Both show an increase in predictability over the last century.

\section{Discussion and conclusion}
\label{sec:conclusion}

In this work, we developed a stochastic thermodynamic framework for overdamped Ornstein-Uhlenbeck (OU) stochastic differential equations and related it to the predictability of climate oscillations using Linear Inverse Modeling (LIM). Using a coarse-grained endpoint formulation together with simplifying coordinate transformations, we showed that the coordinate-invariant properties of the nonequilibrium steady-state of a two-dimensional OU process can be fully characterized by only three independent parameters: the mean diffusion strength $\bar d$, the diffusion anisotropy $\Delta d$, and the probability angular momentum $\ell$. The remaining four parameters describe the observational coordinate system and determine coordinate-dependent properties. The resulting $(\bI,\bLambda_\bD,\bL)$ coordinate system provides a compact representation of the dynamics in terms of diffusion, anisotropy, and rotational probability currents.

Within this framework, endpoint entropy production quantifies the time-directional asymmetry of the coarse-grained dynamics, while endpoint dynamical activity quantifies the overall intensity of fluctuations associated with coarse-grained transitions in phase space independently of temporal directionality. The endpoint action relative to the climatological distribution provides a natural predictability measure, since it quantifies the information gained by using the system dynamics instead of the stationary climatological distribution. In the approximately isotropic regime, the action-based predictability $\rho$ becomes nearly equivalent to the Anomaly Correlation Coefficient (ACC), while remaining coordinate invariant. The decomposition of these quantities also highlights that predictability, irreversibility, and dynamical activity capture distinct aspects of the dynamics. 

The $(\bI,\bLambda_\bD,\bL)$ coordinate system further revealed the existence of distinct dynamical regimes. In the rotational regime, oscillatory behavior arises from the competition between rotational dynamics and diffusion, while in the diffusive regime predictability decays more slowly. In the rotational regime, predictability is controlled primarily by the diffusive timescale, whereas irreversibility depends additionally on rotational strength and anisotropy. Thus, systems with similar predictability can nevertheless exhibit substantially different degrees of irreversibility and dynamical activity.

The application to the Madden–Julian Oscillation suggests that it lies in an approximately isotropic fast-rotational regime — $\Delta d\ll\bar d\ll\ell$ — throughout the twentieth century. When timescales are defined relative to $\ell$ rather than $\bar d$, this regime can be interpreted as a weakly perturbed neutral limit cycle, with small noise and damping and even weaker diffusive anisotropy. In this regime, the long-term evolution of predictability is governed primarily by the mean diffusion strength. The increase in MJO predictability over the twentieth century is therefore associated mainly with a reduction in mean diffusion, while the simultaneous increase in irreversibility and dynamical activity results from the combined evolution of diffusion, anisotropy, and rotational dynamics. The close agreement between the long-term trends of the LIM and CSF-20C predictability estimates further suggests that a substantial fraction of the MJO predictability evolution can be captured by linear Gaussian stochastic dynamics alone.

Several limitations should nevertheless be emphasized. First, within the LIM framework, diffusion represents the aggregate effect of unresolved fast processes, but its temporal evolution may also reflect changes in observational coverage, reanalysis methodology, or data assimilation procedures over the twentieth century. Consequently, the trends identified here should not be interpreted uniquely as changes in the intrinsic physical dynamics of the MJO. Second, the present framework is restricted to linear Gaussian dynamics and a two-dimensional phase-space representation. While these assumptions permit analytical treatment and direct thermodynamic interpretation, real climate variability exhibits important nonlinearities, state dependence, and non-Gaussian features that are not captured by an Ornstein–Uhlenbeck approximation. Nevertheless, the MJO application serves primarily to demonstrate the framework's applicability to real climate data, rather than to provide a definitive physical account of MJO predictability trends. 

Despite these limitations, stochastic thermodynamics provides a useful and analytically tractable framework for interpreting climate predictability from a nonequilibrium perspective. The joint characterization in terms of predictability, irreversibility, and dynamical activity provides a way to distinguish information retention, time-directional asymmetry, and overall fluctuation intensity within a common reduced description. Because two-dimensional OU models are widely used across the physical sciences and can be mapped onto the $(\delta,\lambda)$ parameter space, the framework provides a natural basis for comparing different low-dimensional stochastic systems. An important direction for future work is therefore to apply this approach to other climate oscillations and to investigate whether common dynamical regimes emerge across systems with different physical origins.

An additional advantage of OU dynamics is their analytical tractability and low computational cost, which makes them particularly suitable for ensemble forecasting and for systems where observations are insufficient to support more complex dynamical models. In this sense, OU dynamics can provide a useful null hypothesis for oscillatory behavior: they isolate the contribution of linear stochastic dynamics to predictability and phase-space rotation while providing a baseline against which nonlinear or state-dependent effects can be assessed. More generally, the framework developed here offers a route toward connecting thermodynamic quantities, information-theoretic measures, and conventional predictability metrics across a broad class of low-dimensional stochastic models.

\begin{acknowledgments}
RB has been funded by the University of Bologna under the Future Earth, Climate Change and Societal Challenges PhD Program. JBW was partially supported by NSF DMS 2308338 and by a Bassi Fellowship at the CMCC Foundation (Euro-Mediterranean Center on Climate Change) in Bologna, Italy. We thank Nadia Pinardi and Royce Zia for their encouragement and feedback. 
\end{acknowledgments}

\section*{Data Availability}
CERA-20C and CSF-20C (ECMWF forecasts) data used in this study are publicly available from the dataset associated with Du et al.~(2024) \cite{du2024increase} at \href{https://doi.org/10.6084/m9.figshare.24082053.v1}{doi:10.6084/m9.figshare.24082053.v1} \cite{du2023dataset}. The codes used for the analysis are available upon request.

%\FloatBarrier
%TC:ignore

\appendix

\section{Average thermodynamic quantities of an OU process}
\label{appendix:average_quantities}

From the decomposition of the endpoint probabilities in Eq. \eqref{eq:decomposition_forward_path_prob} and \eqref{eq:decomposition_backward_path_prob} and since the system is Gaussian, we can write the stationary and the transition probability distributions as
\begin{align}
    P_0(\bx) &= \frac{1}{\sqrt{(2\pi)^N \det(\bC_0)}}e^{-\bx^T \bC_0^{-1}\bx/2} \\
    P(\bx_1|\bx_0) &= \frac{1}{\sqrt{(2\pi)^N \det(\bC_t)}}e^{-\Delta \bx^T \bC_t^{-1}\Delta \bx/2}.
\end{align}
Thus, by defining $\bz=\begin{pmatrix}\bx_1\\\bx_2\end{pmatrix}$, the forward and backward endpoint probabilities are
\begin{align}
    P^e(\bx_0,\bx_1) &= \frac{1}{(2\pi)^N\sqrt{ \det(\bC_0 \bC_t)}}e^{-\bz^T \bR_{01}\bz/2} \\
    P^e(\bx_1,\bx_0) &= \frac{1}{(2\pi)^N\sqrt{ \det(\bC_0 \bC_t)}}e^{-\bz^T \bR_{10}\bz/2},
\end{align}
where $\bR_{01}$ and $\bR_{10}$ are $2N \times 2N$ positive definite matrices:
\begin{align}
    \bR_{01} &= \begin{pmatrix}
    e^{\bF^Tt}\bC^{-1}_te^{\bF t} + \bC^{-1}_0& -e^{\bF^Tt}\bC^{-1}_t \\
    -\bC^{-1}_te^{\bF t} & \bC^{-1}_t
    \end{pmatrix} \\
    \bR_{10} &= \begin{pmatrix}
    \bC^{-1}_t & -\bC^{-1}_te^{\bF t} \\
     -e^{\bF^Tt}\bC^{-1}_t& e^{\bF^Tt}\bC^{-1}_te^{\bF t} + \bC^{-1}_0
    \end{pmatrix}.
\end{align}
Then, the endpoint entropy production between $\bx_0$ and $\bx_1$ is given by: 
\begin{equation}
    \sigma^e = \frac{1}{2} \bz^T \bR \bz
\end{equation}
where $\bR = \bR_{10} - \bR_{01}$. The average over the endpoints then reads:
\begin{align}
    \langle \sigma^e \rangle_e &=  \frac{1}{2}  \langle z_i R_{ij} z_j\rangle_e, \nonumber \\
    &= \frac{1}{2}  R_{ij} \langle z_i z_j\rangle_e = \frac{1}{2} R_{ij} (\bR_{01})^{-1}_{ij}, \nonumber \\
    &= \frac{1}{2} \mathrm{tr[\bR \bR_{01}^{-1}]}.
    \label{eq:average_endpoint_ep_1}
\end{align}
and 
\begin{equation}
    \bR_{10}^{-1} = \begin{pmatrix}
    \bC_0 & \bC_0e^{\bF^T t} \\
     e^{\bF t}\bC_0& \bC_0
    \end{pmatrix},
\end{equation}
which is symmetric.
By explicitly computing the trace in Eq. \eqref{eq:average_endpoint_ep_1}, we recover the expression in Eq. \eqref{eq:sigma_endpoint_average}. 

Similarly, endpoint dynamical activity is obtained from its definition in Eq. \eqref{eq:coarse_grained_phi} as
\begin{equation}
    \phi^e = \frac{1}{2} \left[\bz^T \tilde \bR \bz + \ln\det(\bC_0 \bC_t) + 2N \ln(2\pi)\right]
\end{equation}
where $\tilde \bR = \frac{1}{2}(\bR_{10} + \bR_{01})$ and by performing the same calculations, we find for the average over the endpoints:
\begin{align}
    \langle \phi^e \rangle_e &= \frac{1}{2} \left[ \mathrm{tr[\tilde \bR \bR_{01}^{-1}]} + \ln\det(\bC_0 \bC_t)\right] \nonumber \\
    &+ N \ln(2\pi).
\end{align}
and again, by explicitly computing the trace, we get the expression in Eq. \eqref{eq:phi_endpoint_average}.

Thus, $\langle \alpha^e \rangle_e$ reads:
\begin{align}
    \langle \alpha^e \rangle_e &= \langle \phi^e \rangle_e - \frac{1}{2}\langle \sigma^e \rangle_e \nonumber \\
    &= \frac{1}{2}\ln \det(\bC_t \bC_0) + N[1 + \ln(2\pi)]
\end{align}
where we exploit the fact that $\langle \phi^e \rangle_e= \frac{1}{2} \langle \sigma^e \rangle_e + N[1 + \ln(2\pi)] + \frac{1}{2}\ln \det(\bC_0 \bC_t)$. We get the same result by directly computing $\alpha^e$ from its definition. 

For the reference process, instead:
\begin{align}
    \langle \phi^e_0 \rangle_e &=  N \ln(2\pi) + \ln\det(\bC_0) +\nonumber \\
    &\qquad+ \frac{1}{2} \langle \bx_0^T \bC_0^{-1}\bx_0 + \bx_1^T \bC_0^{-1}\bx_1 \rangle_e, \nonumber \\
    &= N \ln(2\pi) + \ln\det(\bC_0) + \nonumber \\
    &\qquad+\frac{1}{2}(\bC_0^{-1})_{ij} \langle x_{0i} x_{0j} +  x_{1i} x_{1j}\rangle_e, \nonumber \\
    &= N \ln(2\pi) + \ln\det(\bC_0) + (\bC_0^{-1})_{ij} (\bC_0)_{ij} \nonumber \\
    &= N \ln(2\pi) + \ln\det(\bC_0) + \mathrm{tr[\bI]} \nonumber \\
    &= N[1 + \ln(2\pi)] + \ln\det(\bC_0).
\end{align}
Thus:
\begin{align}
    \langle \Delta \sigma^e \rangle_e &= \langle \sigma^e \rangle_e,\nonumber \\
    \langle \Delta \phi^e \rangle_e &= \frac{1}{2} \langle \sigma^e \rangle_e + \frac{1}{2}\ln\frac{\det\bC_t}{\det\bC_0},\nonumber \\
    \langle \Delta \alpha^e \rangle_e &= \frac{1}{2}\ln\frac{\det\bC_t}{\det\bC_0}.
\end{align}

\section{\texorpdfstring{Thermodynamics in $(\bI, \mathbf{\Lambda}_\bD, \bL)$ coordinates}
{Thermodynamics in (I, LambdaD, L) coordinates}}
\label{appendix:red_cdl}

Using the reduced CDL coordinate system, the time covariance matrix, according to Eq. \eqref{eq:time_cov_matrix_2}, reads: 
\begin{equation}
    \mathbf{C}_t = \mathbf{I}-e^{\mathbf{F}t}e^{\mathbf{F}^Tt}
    \label{eq:time_cov_cdl_red}
\end{equation}
Thus, we can rewrite Eq. \eqref{eq:sigma_endpoint_average} as:
\begin{equation}
    \langle \sigma^e \rangle_e = \Tr[(\mathbf{I}-e^{\mathbf{F}t}e^{\mathbf{F}^Tt})^{-1}(\mathbf{I}-e^{2\mathbf{F}t})-\mathbf{I}] \\
    \label{eq:sigma_cdl_red}
\end{equation}
Since $[\mathbf{F},\mathbf{F}^T] = -4\ell\Delta d \begin{pmatrix} 0 & 1 \\ 1 & 0\end{pmatrix}$, $\mathbf{F}$ and $\mathbf{F}^T$ commute only if $\ell=0$ or $\Delta d = 0$, corresponding to the equilibrium and isotropic cases, respectively. As a consequence, in general,
\begin{equation}
e^{\mathbf{F} t} e^{\mathbf{F}^T t} \neq e^{(\mathbf{F} + \mathbf{F}^T) t},
\label{eq:exp_not_commuting}
\end{equation}
and the two matrix exponentials must therefore be evaluated separately. In order to compute the exponential, we exploit the Caley-Hamilton theorem for $2 \times 2$ matrix exponentials which reads 
\begin{equation}
    e^{\mathbf{F}t} = \beta_0\mathbf{I}+\beta_1\mathbf{F},
    \label{eq:ch_theorem}
\end{equation}
and, similarly, for the eigenvalues:
\begin{equation}
    e^{\mu_{\pm}t} = \beta_0+\beta_1\mu_{\pm}.
    \label{eq:ch_theorem_eigvals}
\end{equation}
where $\mu_{\pm} = -\bar d \,\pm \, i\omega.$
We choose $\beta_0$ and $\beta_1$ as those that satisfy Eq. \eqref{eq:ch_theorem_eigvals} for both $\mu_{\pm}$, obtaining that:
\begin{equation}
    e^{\mathbf{F}t} = \frac{e^{\mu_{-}t}(\mu_+\mathbf{I}-\mathbf{F})+e^{\mu_{+}t}(\mathbf{F}-\mu_-\mathbf{I})}{\mu_+-\mu_-}.
    \label{eq:exp_F}
\end{equation}
From Eq.~\eqref{eq:exp_F}, the expressions for $e^{\mathbf{F}^T t}$ and $e^{2 \mathbf{F} t}$ follow immediately. The full expression~\eqref{eq:sigma_cdl_red} for $\langle \sigma^e \rangle_e$ can then be evaluated through tedious but straightforward algebra from Eq. \eqref{eq:drift_red_cdl_coord}, yielding Eq.~\eqref{eq:reduced_CDL_sigma}. Similarly, the same decomposition for $\bC_t$ from Eq. \eqref{eq:time_cov_cdl_red} can be used to compute $\langle \Delta \phi^e \rangle_e$ and $\langle \Delta \alpha^e \rangle_e$ explicitly resulting in Eq. \eqref{eq:reduced_CDL_phi} and Eq. \eqref{eq:reduced_CDL_alpha}, respectively.

\begin{widetext}
\section{Information theory based metrics}
\label{appendix:info_theory}
Using the definition of Shannon entropy and Kullback-Leibler divergence \cite{CoverThomas}, it is straightforward to show the equalities of Eq. \eqref{eq:info_theory_coarse_graining_eqs}:

\begin{align}
    \langle \sigma^e \rangle_e &= \int d\bx_0 \, d\bx_1 \, P^e(\bx_0,\bx_1) \ln\frac{P^e(\bx_0, \bx_1)}{P^e(\bx_1, \bx_0)} = \mathcal{D}_{KL}(P^e\|\hat P^e) \nonumber \\
    \langle \phi^e \rangle_e &= -\frac{1}{2}\int d\bx_0 \, d\bx_1 \, P^e(\bx_0,\bx_1) \ln\left[P^e(\bx_0, \bx_1)P^e(\bx_1, \bx_0) \right] \nonumber \\
    &= - \int d\bx_0 \, d\bx_1 \, P^e(\bx_0,\bx_1) \ln P^e(\bx_0, \bx_1) + \frac{1}{2}\int d\bx_0 \, d\bx_1 \, P^e(\bx_0,\bx_1) \ln\frac{P^e(\bx_0, \bx_1)}{P^e(\bx_1, \bx_0)} \nonumber \\
    &= \mathcal{H}[P^e] + \frac{1}{2} \mathcal{D}_{KL}(P^e\|\hat P^e) \nonumber \\
\end{align}
If we consider $P^e_{ref}$ as a generic pdf then :
\begin{align}
\langle \phi^e_{ref} \rangle_e  &= -\frac{1}{2}\int d\bx_0 \, d\bx_1 \, P^e(\bx_0,\bx_1) \ln\left[P_{ref}^e(\bx_0, \bx_1) \hat P_{ref}^e(\bx_1, \bx_0) \right], \nonumber \\
&= \frac{1}{2} (\mathcal{H}[P^e, P^e_{ref}] + \mathcal{H}[P^e, \hat P^e_{ref}]) = \mathcal{H}[P^e] + \frac{1}{2}(\mathcal{D}_{KL}(P^e\|P^e_{ref})+\mathcal{D}_{KL}(P^e\|\hat P^e_{ref})) \\
\langle \sigma^e_{ref} \rangle_e &= \int d\bx_0 \, d\bx_1 \, P^e(\bx_0,\bx_1) \ln\frac{P_{ref}^e(\bx_0, \bx_1)} {\hat P_{ref}^e(\bx_1, \bx_0)}, \nonumber \\
&= -\mathcal{H}[P^e, P^e_{ref}] + \mathcal{H}[P^e, \hat P^e_{ref}] = \mathcal{D}_{KL}(P^e\|\hat P^e_{ref}) - \mathcal{D}_{KL}(P^e\|P^e_{ref}). \\
\langle \Delta \alpha^e \rangle_e &= -\int d\bx_0 \, d\bx_1 \, P^e(\bx_0,\bx_1) \ln\frac{P^e(\bx_0, \bx_1)}{P^e_{ref}(\bx_0, \bx_1)} = - \mathcal{D}_{KL}(P^e\|P^e_{ref}). \nonumber \\
\end{align}
where we used the fact that $\mathcal{H}[P^e, P^e_{ref}] = \mathcal{H}[P^e] + \mathcal{D}_{KL}(P^e\|P^e_{ref})$ and similarly for $\mathcal{H}[P^e, \hat P^e_{ref}]$.

If we choose $P^e_{ref} = P^e_0 = P_0(\bx_0) P_0(\bx_1)$, then:
\begin{align}
    \langle \phi^e_0 \rangle_e  &= -\frac{1}{2}\int d\bx_0 \, d\bx_1 \, P^e(\bx_0,\bx_1) \ln\left[P_0^e(\bx_0, \bx_1) P_0^e(\bx_1, \bx_0) \right], \nonumber \\
    &= -\int d\bx_0 \, d\bx_1 \, P^e(\bx_0,\bx_1) \ln\left[P_0(\bx_0)P_0(\bx_1) \right], \nonumber \\
    &= -\int d\bx_0 \, d\bx_1 \, P^e(\bx_0,\bx_1) \ln\left[P_0^e(\bx_0, \bx_1) \right], \nonumber \\
    & = \mathcal{H}[P^e, P_0^e],\\
\langle \Delta \phi^e \rangle_e &= \mathcal{H}[P^e] - \mathcal{H}[P^e, P_0^e] + \frac{1}{2} \mathcal{D}_{KL}(P^e\|\hat{P}^e) \\
&= -\mathcal{D}_{KL}(P^e\|P^e_0) + \frac{1}{2} \mathcal{D}_{KL}(P^e\|\hat{P}^e)
\end{align}
and $\langle \sigma^e_0 \rangle_e = 0$.

\section{Endpoint bounds on path-dependent thermodynamic quantities}
\label{appendix:inequalities}

Eq. (\ref{eq:exp_sigma_action_total_vs_endpoints}) can be easily deduced from the definition of $\sigma$ and of $\sigma^e$:
\begin{align}
    \langle e^{-\sigma} \rangle_{\bX^e} &= \int_{\bX^e} \mathcal{D}\bX \, P^{\bX^e}[\bX] \, e^{-\sigma} \nonumber \\
    &=\int_{\bX^e} \mathcal{D}\bX \, P^{\bX^e}[\bX] \,\frac{P^{\hat \bX^e}[\hat \bX] P^e(\bx_1, \bx_0)}{P^{\bX^e}[\bX] P^e(\bx_0, \bx_1)} \nonumber \\
    &=\frac{P^e(\bx_1, \bx_0)}{P^e(\bx_0, \bx_1)} \int_{\bX^e} \mathcal{D}\bX \, P^{\hat \bX^e}[\hat \bX] \nonumber \\
    &= e^{-\sigma^e}
\end{align}
where we exploit the normalization $\int_{\bX^e} \mathcal{D}\bX \, P^{\hat \bX^e}[\hat \bX] = 1$. A similar reasoning can be applied for $\Delta \alpha$ recovering Eq. (\ref{eq:exp_sigma_action_total_vs_endpoints}), whereas it does not work for dynamical activity.

The inequality in Eq. (\ref{eq:all_total_vs_endpoints_1}), instead, is recovered from:
\begin{align}
    \langle \sigma \rangle_{\bX^e} &= \int_{\bX^e} \mathcal{D}\bX \, P^{\bX^e}[\bX] \sigma \nonumber \\
    &= \int_{\bX^e} \mathcal{D}\bX \, P^{\bX^e}[\bX] \ln \frac{P^{\bX^e}[\bX] P^e(\bx_0, \bx_1)}{P^{\hat \bX^e}[\hat \bX]P^e(\bx_1, \bx_0)} \nonumber \\
    &= \ln \frac{P^e(\bx_0, \bx_1)}{P^e(\bx_1, \bx_0)} + \int_{\bX^e} \mathcal{D}\bX \, P^{\bX^e}[\bX] \ln \frac{P^{\bX^e}[\bX] }{P^{\hat \bX^e}[\hat \bX]} \nonumber \\ 
    &= \sigma^e + \mathcal{D}_{KL}(P^{\bX^e}\|P^{\hat \bX^e}) \nonumber \\
    &\geq \sigma^e
\end{align}
where the inequality comes from the fact that $\mathcal{D}_{KL} \geq 0$ always. 
    
Similarly, for $\Delta \alpha$:
\begin{align}
    \langle \Delta \alpha \rangle_{\bX^e} &= \int_{\bX^e} \mathcal{D}\bX \, P^{\bX^e}[\bX] \Delta \alpha \nonumber \\
    &= \int_{\bX^e} \mathcal{D}\bX \, P^{\bX^e}[\bX] \ln \frac{P_{ref}^{\bX^e}[\bX] P^e_{ref}(\bx_0, \bx_1)}{P^{\bX^e}[\bX] P^e(\bx_0, \bx_1)} \nonumber \\
    &= \ln \frac{P^e_{ref}(\bx_0, \bx_1)}{P^e(\bx_0, \bx_1)} + \int_{\bX^e} \mathcal{D}\bX \, P^{\bX^e}[\bX] \ln\frac{P_{ref}^{\bX^e}[\bX]}{P^{\bX^e}[\bX]}  \nonumber \\
    &= \Delta \alpha^e - \mathcal{D}_{KL}(P^{\bX^e}\|P_{ref}^{\bX^e}) \leq \Delta \alpha^e
\end{align}

Lastly, for $\phi$:
\begin{align}
    \langle \phi \rangle_{\bX^e} &= \int_{\bX^e} \mathcal{D}\bX \, P^{\bX^e}[\bX] \phi \nonumber \\
    &= -\frac{1}{2}\int_{\bX^e} \mathcal{D}\bX \, P^{\bX^e}[\bX] \ln \left[P^{\bX^e}[\bX] P^e(\bx_0, \bx_1) P^{\hat \bX^e}[\hat \bX] P^e(\bx_1, \bx_0) \right] \nonumber \\
    &= -\frac{1}{2} \ln \left[P^e(\bx_0, \bx_1)P^e(\bx_1, \bx_0) \right] -\frac{1}{2} \int_{\bX^e} \mathcal{D}\bX \, P^{\bX^e}[\bX] \ln \left[P^{\bX^e}[\bX]  P^{\hat \bX^e}[\hat \bX] \right]  \nonumber \\
    &= \phi^e - \int_{\bX^e} \mathcal{D}\bX \, P^{\bX^e}[\bX]  \ln P^{\bX^e}[\bX] -\frac{1}{2} \int_{\bX^e} \mathcal{D}\bX \, P^{\bX^e}[\bX] \ln \frac{P^{\hat \bX^e}[\hat \bX]}{P^{\bX^e}[\bX]} \nonumber \\
    &= \phi^e +\mathcal{H}(P^{\bX^e}) + \frac{1}{2} \mathcal{D}_{KL}(P^{\bX^e}\|P^{\hat \bX^e}).
\end{align}

Since $\mathcal{H}(P^{\bX^e})$ does not have a definite sign, we cannot write an inequality for $\phi$. The same holds if we consider $\Delta \phi$ instead. We get that $\langle \Delta \phi \rangle_e = \Delta \phi_e$ only when $P^{\bX^e}=P^{\bX^e}_{ref}$. In that case, $\langle \Delta \alpha \rangle_e = \Delta \alpha^e$ and $\langle \Delta \sigma \rangle_e = \Delta \sigma^e$ as well. 

\section{ACC for an OU process}
\label{appendix:acc_langevin}

From the definition of the ACC in Eq. \eqref{eq:acc_langevin}, we can explicitly compute the different terms. Note that we use Einstein notation and that we exploit the fact that $\bC_0$ and $\bC_t$ are symmetric.

\begin{align}
    \mathbb{E}[\bar \bx^T(t)\bar \bx(t)] &= \mathbb{E}[\mathbf{x}^T(t_i)e^{\mathbf{F}^Tt}e^{\mathbf{F}t}\mathbf{x}(t_i)] \nonumber \\
    &= \mathbb{E}\left[x_k e^{\mathbf F^T t}_{kj}e^{\mathbf F t}_{jl}x_l\right] \nonumber \\
    &= e^{\mathbf F^T t}_{kj}e^{\mathbf F t}_{jl} \, \mathbb{E}[x_{k} x_{l}] \nonumber \\
    &= (e^{\mathbf{F}^Tt}e^{\mathbf{F}t})_{kl}[\mathbf{C_{0}}]_{kl} \nonumber \\
    &= \mathrm{tr}[e^{\mathbf{F}t}\mathbf{C}_{0}e^{\mathbf{F}^Tt}]
    \label{C_bb}
\end{align}
\allowdisplaybreaks
where $\mathbb{E}[ \cdot ]$ is the average over the ensemble and over all initializations.
\begin{align}
    \mathbb{E}[\bx^T(t)\bx(t)] 
    &= \mathbb{E}[\mathbf{x_{n}}^T(t;t_i)\mathbf{x_{n}}(t;t_i)]  \nonumber \\
    &= \mathbb{E}[\bar \bx^T(t)\bar \bx(t)] + \mathbb{E}\left[\int^t_0 ds_1 \int^t_0 ds_2 \boldsymbol{\xi}_{n}^T(s_1)\sqrt{2\mathbf{D}^T}e^{\mathbf{F}^T(t - s_1)}e^{\mathbf{F}(t - s_2)} \sqrt{2\mathbf{D}}\boldsymbol{\xi}_{n}(s_2)\right] \nonumber \\
    &= \mathrm{tr}[e^{\mathbf{F}t}\mathbf{C}_{0}e^{\mathbf{F}^Tt}] +\mathbb{E}\left[\int^t_0 ds_1 \int^t_0 ds_2 \xi_{j}(s_1)\sqrt{2D_{kj}}e_{lk}^{\mathbf{F}^T(t - s_1)}e_{lm}^{\mathbf{F}(t - s_2)}\sqrt{2D_{mi}}\boldsymbol{\xi}_{i}(s_2)\right]  \nonumber \\
    &= \mathrm{tr}[e^{\mathbf{F}t}\mathbf{C}_{0}e^{\mathbf{F}^Tt}] + \int^t_0 ds_1 \int^t_0 ds_2 \, \mathbb{E}[\xi_{j}(s_1)\xi_{i}(s_2)] \,\sqrt{2D_{kj}} e_{lk}^{\mathbf{F}^T(t - s_1)} e_{lm}^{\mathbf{F}(t - s_2)}\sqrt{2D_{mi}} \nonumber \\
    &= \mathrm{tr}[e^{\mathbf{F}t}\mathbf{C}_{0}e^{\mathbf{F}^Tt}] + \int^t ds e_{lk}^{\mathbf{F}^T(t - s)}[\sqrt{2\bD}\sqrt{2\bD^T}]_{km}e_{lm}^{\mathbf{F}(t - s)} \nonumber \\
    &= \mathrm{tr}[e^{\mathbf{F}t}\mathbf{C}_{0}e^{\mathbf{F}^Tt}] + 2\int^t_0 ds e_{lk}^{\mathbf{F}^T(t - s)}D_{km}e_{lm}^{\mathbf{F}(t - s)} \nonumber \\
    &= \mathrm{tr}[e^{\mathbf{F}t}\mathbf{C}_{0}e^{\mathbf{F}^Tt}] + 2 \mathrm{tr}\left[\int^t_0 ds e^{\mathbf{F}^T(t - s)}\mathbf{D}e^{\mathbf{F}(t - s)}\right] \nonumber \\
    &= \mathrm{tr}[e^{\mathbf{F}t}\mathbf{C}_{0}e^{\mathbf{F}^Tt} + \mathbf{C}_{t}] = \mathrm{tr}[\mathbf{C}_{0}]
    \label{C_aa}
\end{align}
and
\begin{eqnarray}
    \mathbb{E}[\bx^T(t)\bar \bx(t)] &= \mathbb{E}[\bar \bx^T(t)\bar \bx(t)] = \mathrm{tr}[\mathbf{C}_{0} - \mathbf{C}_{t}]
    \label{C_ab}
\end{eqnarray}
where we used the fact that $\mathbb{E}[\xi^n_i(s)\xi^m_j(t)] = \delta_{n,m} \delta_{i,j}\delta(t-s)$, 
$\mathbb{E}[\xi^n_i] = 0$ and Eq. \eqref{eq:time_cov_matrix}. \\
The ACC for a Langevin system then reduces to Eq. \eqref{eq:acc_langevin}:
\begin{equation}
    ACC = \sqrt{\frac{\mathrm{tr}[\bC_0 - \bC_t]}{\mathrm{tr}[\bC_0]}} \nonumber
\end{equation}
By exploiting the same relations found in Appendix \ref{appendix:red_cdl},  in $(\bI, \mathbf{\Lambda}_\bD, \bL)$ coordinates Eq. \eqref{eq:acc_langevin} reads:
\begin{equation}
    ACC_t = \sqrt{\frac{\mathrm{tr}[e^{\mathbf{F}t}e^{\mathbf{F}^Tt}]}{2}}  = e^{-\bar d t}
    \sqrt{\frac{\ell^2 - \Delta d^2\cos(2t\sqrt{\ell^2 - \Delta d^2})}{\ell^2 - \Delta d^2}}. \nonumber
\end{equation}
which yields Eq. \eqref{eq:acc_CDL} when scaled by $\bar d$.
\end{widetext}
\section{Potential predictability measures}
\label{appendix:skill_scores}

We measure the potential predictability captured by each model by comparing the forecast ensemble mean with each ensemble member taken alternately as truth using a correlation score, a phase error, an amplitude error and the Root Mean Square Error (RMSE):
    \begin{align}
    ACC(t,y) &= \sqrt{\frac{\langle\mathbf{a}^T\mathbf{b}\rangle_{N_i, N_e, w}^2}{\langle\mathbf{a}^T\mathbf{a}\rangle_{N_i, N_e, w}\langle\mathbf{b}^T\mathbf{b}\rangle_{N_i, N_e, w}}}\\
    \Delta \theta(t,y) &= \langle\arctan{\frac{\mathbf{ab}^T}{\mathbf{a}^T\mathbf{b}}}\rangle_{N_i, N_e, w}\\ 
    \Delta A(t,y) &= \langle\sqrt{\mathbf{a}^T\mathbf{a}}-\sqrt{\mathbf{b}^T\mathbf{b}}\rangle_{N_i, N_e, w} \\
    RMSE(t,y) &= \sqrt{\langle(\mathbf{a}-\mathbf{b})^T(\mathbf{a}-\mathbf{b})\rangle_{N_i, N_e, w}} 
    \label{eq:acc_danni}
    \end{align}
where we defined the mean operator as $\langle \, \cdot \,\rangle_{N_i, N_e, w} = \frac{1}{\sum_{i=1}^{N_i} N_e^i} \sum^{N_i}_{i = 1} \frac{1}{N_e^i}\sum^{N_e^i}_{e = 1} \frac{1}{w}\sum^{y+w}_{j=y-w}$: $i$ denotes the initialization, $e$ the ensemble member, $t$ the lead time, $y$ the year over which a centered 30-year mean ($w = 30$) is computed. Here, $\mathbf{a}$ is the $e^{\mathrm{th}}$ member of the ensemble of the $i^{\mathrm{th}}$ initialization and $\mathbf{b} = \frac{1}{N_e^i -1}\sum^{N_e^i}_{e' \neq e} \mathbf{a}$ is the average over the ensemble excluding the $e^{\mathrm{th}}$ member. Following standard practice, the phase error is computed only for amplitudes greater than one at $t=0$. The dependence of the ensemble size $N_e^i$ on $i$ takes into account the possibility that the ensemble size could depend on the initialization, which is indeed our case with the CSF-20C dataset. 

Note that these measures are equivalent to those used for the forecast skill assessment taken from \cite{rashid2011prediction} using, instead of the reanalysis, each ensemble member and then averaging over the ensemble. 

In addition, the RMSE of the LIM can be simplified by adopting the notation of Appendix \ref{appendix:acc_langevin} for $\mathbf{a}$ and $\mathbf{b}$ and the result for the ACC of a Langevin system:
\begin{align}
    RMSE_t &= \sqrt{\langle(\mathbf{a}-\mathbf{b})^T(\mathbf{a}-\mathbf{b})\rangle} \nonumber \\
    &= \sqrt{\langle \mathbf{a}^T\mathbf{a} \rangle+\langle \mathbf{b}^T\mathbf{b} \rangle -2\langle \mathbf{a}^T\mathbf{b} \rangle} \nonumber \\
    &=\sqrt{\mathbb{E}[\bar \bx^T\bar \bx+\bx^T\bx -2\bx^T\bar \bx]} \nonumber \\ 
    &= \sqrt{\mathrm{tr}[\mathbf{C}_0]- \mathrm{tr}[\mathbf{C}_0 - \mathbf{C}_t]} = \sqrt{\mathrm{tr}[\mathbf{C}_t]} 
\end{align}
This expression is not coordinate invariant, and by normalizing it by $\sqrt{\mathrm{tr}[\mathbf{C}_0]}$, we get the same information as the ACC:  $ACC =\sqrt{1 - RMSE^2}$.

\bibliographystyle{ieeetr}
\bibliography{references}

\end{document}